\documentclass[12pt,onecolumn,oneside,notitlepage,final]{article}
\usepackage{amsmath}
\usepackage{amssymb}
\usepackage{amsthm}
\usepackage{amscd}
\usepackage{color}
\usepackage{enumerate}
\usepackage{a4wide}
\usepackage{graphics}
\usepackage{hyperref}
\usepackage{times,txfonts}

\newtheorem{theorem}{Theorem}[section] 

\newtheorem{proposition}[theorem]{Proposition}
\newtheorem{lemma}[theorem]{Lemma}
\newtheorem{corollary}[theorem]{Corollary}
\newtheorem{definition}[theorem]{Definition}
\newcommand{\PP}{\mathcal{P}}

\title{$A_{k}$ Generalization of the $O(1)$ Loop Model on a Cylinder: Affine Hecke Algebra, 
$q$-KZ Equation and the Sum Rule}
\author{
Keiichi Shigechi\footnote{E-mail: shigechi@monet.phys.s.u-tokyo.ac.jp} and 
Masaru Uchiyama\footnote{E-mail: uchiyama@monet.phys.s.u-tokyo.ac.jp} \\[0.5cm]
{\it Department of Physics, Graduate School of Science, University of Tokyo}\\
{\it  7-3-1 Hongo, Bunkyo-ku, Tokyo 113-0033, Japan}}
\date{March 26, 2007}

\begin{document}

\maketitle
\begin{abstract}
We study the $A_{k}$ generalized model of the $O(1)$ loop model on a cylinder. 
The affine Hecke algebra associated with the model is characterized by a 
vanishing condition, the cylindric relation. 
We present two representations of the algebra: the first one is the spin 
representation, and the other is in the vector space of states of the $A_{k}$ generalized model. 
A state of the model is a natural generalization of a link pattern.
We propose a new graphical way of dealing with the Yang-Baxter equation 
and $q$-symmetrizers by the use of the rhombus tiling. 
The relation between two representations and the meaning of the 
cylindric relations are clarified through the rhombus tiling. 
The sum rule for this model is obtained by solving the $q$-KZ 
equation at the Razumov-Stroganov point. 
\end{abstract}

PACS: 02.20.Uw, 02.30.Ik, 

short title: $A_{k}$ Generalization of the $O(1)$ Loop Model on a Cylinder

\newpage
\setcounter{tocdepth}{2}
\tableofcontents

\section{Introduction}
The ground state of the $O(1)$ loop model (or the Temperley-Lieb (TL) stochastic
process) has been extensively studied since the observation by Razumov and Stroganov~\cite{RazStr00} (see 
also~\cite{RazStr01a, RazStr01b,RazStr01c,BatdeGNie01,MitNiedeGieBat04,PearRitdeGierNie02,PearRitdeGier01}). 
Through those studies, different research areas in mathematics and physics make contact each other; for 
example, alternating sign matrices (ASMs) in combinatorics~\cite{Zeil96,Kup96} (see~\cite{Bre99} and references therein),
polynomial representations of  the Temperley-Lieb/Hecke algebra in representation theory~\cite{Mart93,Pas05,KasPas06,KasTak06a}, 
and exactly solvable models such as the six-vertex model in statistical physics (for example~\cite{Bax82,WadDegAku89}).

Razumov and Stroganov submitted seven conjectures related to the $XXZ$ spin chain model at the 
anisotropic parameter $\Delta=-1/2$ with periodic conditions in~\cite{RazStr00}. These conjectures were generalized to 
the $O(1)$ loop model in~\cite{BatdeGNie01}. A typical one is the sum rule (Conjecture 8 in~\cite{BatdeGNie01}): 
the $1$-sum of the ground state wavefunction of the $O(1)$ loop model with periodic boundary conditions 
and length $L=2n$ is given by the total number of $n\times n$ ASMs. The ground state wavefunction $\Psi$ has another 
remarkable property. In~\cite{RazStr01b}, it is conjectured that an entry $\Psi_{\pi}$ of $\Psi$ is equal to the total 
number of fully packed loops with a link pattern $\pi$. This is the most popular form of the Razumov-Stroganov (RS) 
conjectures. The above mentioned sum rule is a consequence of the RS conjectures for entries.

The sum rule for the $O(1)$ loop model with periodic boundary conditions was proved by Di~Francesco and Zinn-Justin 
by introducing inhomogeneity and utilizing the integrability~\cite{DiFZJ04a}. 
The key is the $q$-Kniznik-\! Zamoldchikov ($q$-KZ) equation~\cite{FreResh92}, which is equivalent to finding 
the eigenvector of the transfer matrix with eigenvalue unity. At the RS point, {\it i.e.} $q=-\exp(\pi i/3)$, 
by solving the $q$-KZ equation it was found that the sum of the entries of $\Psi$ is equal to the partition 
function of the six-vertex model with domain wall boundary conditions (6V/DWBC). If we take the homogeneous 
limit where all the inhomogeneous spectral parameters $z_{i}$'s tend to unity, the partition function of 
6V/DWBC is proportional to the total number of ASMs~\cite{Kup96,Kup02}. 

There are two natural generalizations of the $O(1)$ loop model. The first one is to change 
the geometry of the states, or equivalently to change the boundary conditions ~\cite{DiFZJ06d}. 
The TL algebra naturally acts on the space of link patterns. 
The link patterns considered in~\cite{DiFZJ04a} are undirected ones. On the other hand, we may impose the direction on 
link patterns like those in~\cite{DiFZJZu06}. The former ones are the same as link patterns with periodic 
boundary conditions or on an unpunctured disc, whereas the latter ones are the same as link patterns with cylindric 
boundary conditions or on a punctured disc, {\it i.e.} on a cylinder. 
The direction of link detects the position of the punctured 
point. The space of the link patterns with periodic (resp. cylindric) boundary conditions is 
also equivalent to the space of Dyck paths or restricted (resp. unrestricted) paths of the IRF model.  
The second is to extend the affine TL algebra to the affine Hecke algebra. 
The $A_{k}$ generalized model defined in terms of the Hecke algebra of type $A$~\cite{DiFZJ05a} is a natural 
generalization of the $O(1)$ loop model with periodic boundary conditions. 
Higher-rank models with open boundaries were also discussed in~\cite{DiF05b}.

Remarkably, the eigenvector of the transfer matrix of the $O(1)$ loop model with periodic boundary conditions 
constitutes a special polynomial representation of the affine TL algebra~\cite{Pas05}. This correspondence 
was also studied for the $O(1)$ model with cylindric boundary conditions in~\cite{KasPas06}. 
On the other hand, the special polynomial solutions of $q$-KZ equation for the higher-rank case of 
$U_{q}(\widehat{sl_{k}})$ was constructed in~\cite{KasTak06a}. 

In this paper, we define and study the $A_{k}$ generalized model on a cylinder. 
This model is a new hybrid generalization of the $O(1)$ loop model; defined in terms of the affine 
Hecke algebra of type $A$ and with cylindric boundary conditions. The affine Hecke algebra satisfies new 
vanishing conditions, which we call ``the cylindric relations" 
(see Eqn.(\ref{cyclic-1}) and (\ref{cyclic-general}) in Section~\ref{sec-Hecke-def}). 
The cylindric relations fix the spin representation of the affine Hecke 
algebra. The intuitive meaning of the cylindric relation is to assign to a ``band" around the 
cylinder a certain weight written in terms of the second kind of the Chebyshev polynomials. 
This is a natural generalization for the affine TL algebra considered in~\cite{DiFZJZu06}, where the weight of a 
loop around the cylinder is $\tau$. 
We establish an explicit way of constructing states of the $A_{k}$ generalized model. 
Each state is written in terms of the affine Hecke algebra and characterized by a path. 
For this purpose, we introduce a novel graphical way of depicting states by the use of the rhombus tiling. 
Although similar graphs appeared in the IRF model and the paper~\cite{DiFZJ05a}, our graphical way has the 
following properties. A rhombus represents the $\check{R}$-matrix constructed from the affine Hecke generator. 
On its face, a rhombus has a positive integer indicating the spectral parameter of the $\check{R}$-matrix. 
The Yang-Baxter equation is realized as the equivalence between two different ways of tiling of a hexagon. 
The $q$-symmetrizer $Y_{k}$ of the affine Hecke algebra is expressed 
as a $2(k+1)$-gon. A state is identified with a path  via the graphical representation of the rhombus tiling. 
Roughly speaking, piling rhombus tiles over the $2(k+1)$-gons and reading the path on the top of 
the rhombus tiling, we have an unrestricted path. Indeed, an unrestricted path gives a representation of a state
of our model.  

We consider the eigenvector of the transfer matrix of the $A_{k}$ generalized model with eigenvalue unity  
at the Razumov-Stroganov point, $q=-\exp(\pi i/(k+1))$. At this point, the eigenvector of the transfer matrix 
satisfies the $q$-KZ equation. Originally, the $q$-KZ equation has two parameters $q$ and $s$. The parameter $s$
indicates the action of the cyclic transformation. In our situation, however, the definition of the $A_{k}$ 
generalized model has only $q$ and assumes $s=1$. Due to the condition $s=1$, the eigenvector of the 
transfer matrix with eigenvalue unity should coincide with the solution of the $q$-KZ equation only at the 
RS point.

By resolving the solution of the $q$-KZ equation at the RS point, we find that the 
sum of the weighted entries is the product of $k$ Schur functions. 
Our sum rule contains the sum rule for the $O(1)$ loop model on a cylinder when $k=2$~\cite{DiFZJZu06}. 
Compared with  the results in~\cite{KasTak06a} (\cite{KasPas06} for $k=2$), the solution of the $q$-KZ equation 
obtained in this paper is identified with the one of level $1+\frac{1}{k}-k$ . 

This paper is organized as follows. In Section~\ref{sec-Hecke}, we briefly review the 
affine Hecke algebra. We introduce a class of the affine Hecke algebra which 
is characterized by the cylindric relation. 
The graphical definition and some basic properties of a rhombus with an integer are given. 
In Section~\ref{sec-Spin}, we consider 
the spin representation of the affine Hecke algebra. We show that the affine Hecke generator is obtained by 
twisting the standard Hecke generator by a diagonal matrix. Most parts of Section~\ref{sec-Spin} 
are devoted to the proof of the cylindric relations in the spin representation. 
In Section~\ref{sec-Akmodel}, we move to the $A_{k}$ generalized model on a cylinder. 
We first briefly introduce the $O(1)$ loop model on a cylinder with the perimeter of even 
length and reproduce the sum rule in Section~\ref{Sec-qKZ}. 
The derivation of the sum rule is different from~\cite{DiFZJZu06} in the sense that we consider only the even 
case. We consider the space of link patterns which the affine 
Temperley-Lieb algebra acts on. We also explicitly write down the word representation of  
the highest weight state. Then, we obtain the sum rule for the $O(1)$ loop model by solving 
the $q$-KZ equation. 
In Section~\ref{subsec-Akmodel}, we introduce the $A_{k}$ generalization of the $O(1)$ loop
model on a cylinder. We construct the states for this model through the correspondence among 
an unrestricted path, a rhombus tiling and a word. The relation to the 
spin chain model is also stated in Section~\ref{subsec-Ak-spin}. We solve the $q$-KZ equation and obtain the sum rule 
in Section~\ref{subsec-qKZAk}. This solution is identified with the solution of the $q$-KZ equation of 
level $1+\frac{1}{k}-k$ in Section~\ref{subsec-specialqKZ}.
Section~\ref{sec-schur} is devoted to the evaluation of the recursive relation for the 
Schur function appeared in Section~\ref{sec-Akmodel}. Concluding remarks are in Section~\ref{sec-conclusion}. 
In Appendix~\ref{Cipipi'}, we show that some coefficients $C_{i,\pi,\pi^{'}}$ (see Section~\ref{sec-Akmodel}) 
are equal to 1. In Appendix~\ref{examples-rep}, we give examples how the affine Hecke algebra acts on a 
state in the case of $(k,n)=(3,1)$ and $k=4$.

\section{Affine Hecke Algebra}\label{sec-Hecke}
We introduce the affine Hecke algebra of type $A$ in Section~\ref{sec-Hecke-def}. We impose a new vanishing 
condition on the affine Hecke algebra, which we call the ``cylindric relations". In Section~\ref{sec-Hecke-qsym} 
and \ref{sec-Hecke-YB}, we present basic properties of $q$-symmetrizers and the Yang-Baxter equation. 
In Section~\ref{subsec-graph-Y}, a rhombus with an integer is introduced for a new graphical method. The graphical ways 
of the Yang-Baxter equation and $q$-symmetrizers are given by rhombus tiling.

\subsection{Affine Hecke algebra}\label{sec-Hecke-def}
The Iwahori-Hecke algebra $H_{N}(\tau)$ has generators $\{e_{1},\cdots,e_{N-1}\}$ 
which satisfy the following defining relations:
\begin{subequations}\label{Hecke-relation}
\begin{eqnarray}
e_{i}^{2}&=&\tau e_{i}, \\
e_{i}e_{i\pm1}e_{i}-e_{i}&=&e_{i\pm1}e_{i}e_{i\pm1}-e_{i\pm1}, \\
e_{i}e_{j}&=&e_{j}e_{i}, \ \ \ \mathrm{if}\ |i-j|>1, 
\end{eqnarray}
\end{subequations}
where we set $\tau=-(q+q^{-1})$. If we set $t_{i}=e_{i}+q$, the Iwahori-Hecke algebra can be regarded as
the quotient algebra of the braid group: $t_{i}t_{j}=t_{j}t_{i}$ for $|i-j|>1$, 
$t_{i}t_{i\pm1}t_{i}=t_{i\pm1}t_{i}t_{i\pm1}$ and the quotient relation $(t_{i}-q)(t_{i}+q^{-1})=0$.

When the algebra $H_{N}(\tau)$ satisfies the vanishing condition 
\begin{eqnarray}\label{vc-Hecke}
Y_{k}(e_{i},\cdots,e_{i+k-1})=0, \ \ \mathrm{for}\ i=1,\cdots,N-k,
\end{eqnarray}
we denote this $U_{q}(\mathfrak{sl}(k))$ quotient Hecke algebra by $H_{N}^{(k)}(\tau)$. 
Here, the relation $Y_{m}$ is the Young's $q$-symmetrizer, defined recursively as
\begin{eqnarray}\label{q-Sym-rr}
Y_{m+1}(e_{i},\cdots,e_{i+m})=Y_{m}(e_{i},\cdots,e_{i+m-1})(e_{i+m}-\mu_{m})Y_{m}(e_{i},\cdots,e_{i+m-1})
\end{eqnarray}
with $Y_{1}(e_{i})=e_{i}$ where $\mu_{m}=\frac{U_{m-1}(\tau)}{U_{m}(\tau)}$ and $U_{m}:=U_{m}(\tau)$ is the Chebyshev 
polynomials of the second kind subject to $U_{m}(2\cos x)=\frac{\sin(m+1)x}{\sin x}$. 
The explicit expression of $U_{m}$ is given by
\begin{eqnarray}
U_{m}(\tau)=(-)^{m}\frac{q^{m+1}-q^{-(m+1)}}{q-q^{-1}}, \\
\mu_{m}=-\frac{q^{m}-q^{-m}}{q^{m+1}-q^{-(m+1)}}.
\end{eqnarray}
In particular, $H_{N}^{(2)}(\tau)$ is the Temperley-Lieb algebra. 
\bigskip

The affine Hecke algebra is an extension of the Iwahori-Hecke algebra, obtained by adding the generators 
$y_{i},1\le i\le N$. The generators satisfy (\ref{Hecke-relation}) and 
\begin{eqnarray}
y_{i}y_{j}&=&y_{j}y_{i} \\
t_{i}y_{j}&=&y_{j}t_{i}\qquad \mathrm{if}\ j\neq i,i+1 \\
t_{i}y_{i+1}&=&y_{i}t_{i}^{-1}\qquad \mathrm{if}\ i\le N-1.
\end{eqnarray}

Let us introduce the cyclic operator $\sigma$ through Yang's realization of the affine relation. Then, 
$y_{n}$ is obtained from the recursive relation $y_{n}=t_{n-1}^{-1}y_{n-1}t_{n-1}^{-1}$ with 
$y_{1}=t_{1}t_{2}\cdots t_{n-1}\sigma$.

We may define an additional generator $t_{N}=\sigma t_{1}\sigma^{-1}$, or $e_{N}=\sigma e_{1}\sigma^{-1}$.  
Note that the cyclic operator $\sigma$ makes the defining relations (\ref{Hecke-relation}) 
become cyclic and it holds the relations $\sigma t_{i}=t_{i-1}\sigma$ for all $i$. 
In what follows, we mainly focus on the generators  $\{e_{1},\ldots,e_{N},\sigma\}$, since the affine 
Hecke algebra can be constructed from these generators.

\bigskip

In this paper, we consider the case where $N$ is a multiple of $k$, {\it i.e.}, $N=nk$ with an positive integer 
$n$ and also consider the special case of the affine Hecke algebra $\widehat{H_{N}^{(k)}}(\tau)$ 
by imposing the additional vanishing condition as follows. 
\begin{itemize}
  \item When $N=k$, we have 
\begin{eqnarray}\label{cyclic-1}
Y_{k-1}(e_{1},\cdots,e_{N-1})(e_{N}-\tau)Y_{k-1}(e_{1},\cdots,e_{N-1})=0.
\end{eqnarray}
Obviously, $Y_{k}(e_{1},\cdots,e_{N})$ is non-zero.

  \item  For $N=nk$ with $n\ge2$
\begin{eqnarray}\label{cyclic-general}
Y_{q\text{-sym}}\cdot\prod_{i=1}^{n-1}(e_{ik}-\mu_{k-1})(e_{nk}-\tau)\cdot Y_{q\text{-sym}}=0
\end{eqnarray}
where $Y_{q\text{-sym}}:=\prod_{i=0}^{n-1}Y_{k-1}(e_{ik+1},\cdots,e_{(i+1)k-1})$ is the product of the 
$q$-symmetrizers.
\end{itemize}
Below we call these vanishing conditions as the {\it cylindric relations}. 

\bigskip
\noindent
{\bf Remark1} When $n\ge 2$, the quotient relation (\ref{vc-Hecke}) also becomes cyclic. When $n=1$, the 
cylindric relation can be regarded as a modified quotient relation. The reason why the vanishing condition 
(\ref{vc-Hecke}) breaks will become clear when we consider the spin representation in Section 2.

\bigskip
\noindent
{\bf Remark2} If we set $\sigma=t_{N-1}^{-1}\cdots t_{1}^{-1}$, we obtain the affine Hecke algebra considered 
in \cite{Pas05,DiFZJZu06}.

\subsection{Basic properties of $q$-symmetrizer}\label{sec-Hecke-qsym}
For later convenience, we abbreviate $Y_{k}(e_{i},\cdots,e_{i+k-1})$ as $Y_{k}^{(i)}$. Then, 
we have
\begin{proposition}\label{prop-qY}
The $q$-symmetrizer satisfies the following properties:
\begin{itemize}
 \item $e_{j}Y_{k}^{(i)}=Y_{k}^{(i)}e_{j}=\tau Y_{k}^{(i)}$ if $i\le j\le i+k-1$,
 \item When $l\le k$ and $i\le j\le i+k-l$, we have $Y_{k}^{(i)}Y_{l}^{(j)}=Y_{l}^{(j)}Y_{k}^{(i)}=\alpha_{l}Y_{k}^{(i)}$ 
where $\alpha_{l}=\prod_{i=1}^{k}\mu_{i}^{-2^{k-i}}$.
\end{itemize}
\end{proposition}
\begin{proof} We use the method of induction. When $k=1$, we have 
$e_{i}Y_{1}^{(i)}=Y_{1}^{(i)}e_{i}=e_{i}^{2}=\tau e_{i}$
from the definition of $Y_{1}^{(i)}$. We assume that the statement holds true for less than or equal to $n$, 
{\textit i.e.} $e_{j}Y_{n}^{(i)}=\tau Y_{n}^{(i)}$ for $i\le j\le i+n-1$. From 
the definition $Y_{n+1}^{(i)}=Y_{n}^{(i)}(e_{i+n}-\mu_{n})Y_{n}^{(i)}$, we have 
$e_{j}Y_{n+1}^{(i)}=e_{j}Y_{n}^{(i)}(e_{i+n}-\mu_{n})Y_{n}^{(i)}=\tau Y_{n+1}^{(i)}$ for $i\le j\le i+n-1$. 
Since $Y_{l}^{(j)}$ consists of the generators $e_{i},\cdots,e_{i+l-1}$, 
we also have $Y_{l}^{(j)}Y_{n}^{(i)}=\alpha_{l}Y_{n}^{(i)}$ for $l\le n$ and $i\le j\le i+n-l$. 
Then, the action of $e_{i+n}$ on $Y_{n+1}^{(i)}$ is calculated as
\begin{eqnarray*}
e_{i+n}Y_{n+1}^{(i)}&=&e_{i+n}Y_{n-1}^{(i)}(e_{i+n-1}-\mu_{n-1})Y_{n-1}^{(i)}(e_{i+n}-\mu_{n})Y_{n}^{(i)} \\
&=&Y_{n-1}^{(i)}e_{i+n}(e_{i+n-1}-\mu_{n-1})(e_{i+n}-\mu_{n})Y_{n-1}^{(i)}Y_{n}^{(i)} \\
&=&\alpha_{n-1}Y_{n-1}^{(i)}e_{i+n}(e_{i+n-1}-\mu_{n-1})(e_{i+n}-\mu_{n})Y_{n}^{(i)} \\
&=&\tau\alpha_{n-1}Y_{n-1}^{(i)}(e_{i+n-1}-\mu_{n-1})(e_{i+n}-\mu_{n})Y_{n}^{(i)} \\
&=&\tau Y_{n+1}^{(i)}
\end{eqnarray*}
where we have used in the last equality the relations $e_{i+n-1}Y_{n}^{(i)}=\tau Y_{n}^{(i)}$, 
$e_{i}e_{i+1}e_{i}-e_{i}=e_{i+1}e_{i}e_{i+1}-e_{i+1}$ and $\tau\mu_{j}-\mu_{j}\mu_{j-1}=1$ for any $j$. 
We also have $Y_{n}^{(i)}e_{j}=\tau Y_{n}^{(i)}$ in the similar way. The proof of the first property is then completed. 

We consider the case $Y_{n}=Y_{n}^{(1)}$ for the second relation. We have 
$Y_{k}\cdot Y_{k}=\alpha_{k}Y_{k}$ from the definition of $\alpha_{k}$. On the other hand, by using 
inductive relations, we have
\begin{eqnarray*}
Y_{k}\cdot Y_{k}&=&Y_{k-1}(e_{k-1}-\mu_{k-1})Y_{k-1}Y_{k-1}(e_{k-1}-\mu_{k-1})Y_{k-1} \\
&=&\alpha_{k-1}Y_{k-1}(e_{k}-\mu_{k-1})Y_{k-2}(e_{k-1}-\mu_{k-2})Y_{k-2}(e_{k}-\mu_{k-1})Y_{k-1} \\
&=&\alpha_{k-1}\alpha_{k-2}^{2}Y_{k-1}(e_{k}-\mu_{k-1})(e_{k-1}-\mu_{k-2})(e_{k}-\mu_{k-1})Y_{k-1} \\
&=&\alpha_{k-1}\alpha_{k-2}^{2}\frac{1}{\mu_{k}\mu_{k-1}}Y_{k}. 
\end{eqnarray*}
The proof is completed by checking $\alpha_{n}$ the recurrence relation,
\begin{eqnarray}\label{recur-alpha}
\alpha_{n}=\frac{1}{\mu_{n}\mu_{n-1}}\alpha_{n-1}\alpha_{n-2}^{2},
\end{eqnarray}
with the initial condition $\alpha_{1}=\tau=\mu_{1}^{-1}$ and 
$\alpha_{2}=\tau(\tau^{2}-1)=\mu_{1}^{-2}\mu_{2}^{-1}$, is satisfied by 
\begin{eqnarray}\label{eigenvalue-Y}
\alpha_{n}=\prod_{i=1}^{n}\mu_{i}^{-2^{n-i}}.
\end{eqnarray}
\end{proof}

\subsection{$\check{R}$-matrix and the Yang-Baxter equation}\label{sec-Hecke-YB}
\subsubsection{$\check{R}$-matrix and the Yang-Baxter equation}
Let $\check{R}_{ii+1}=e_{i}+q$ be the $\check{R}$-matrix where $e_{i}$ is the generator of $\widehat{H_{N}^{(k)}}$. 
One can show that $\check{R}_{ii+1}$ satisfies the braid relation
\begin{eqnarray}
\check{R}_{ii+1}\check{R}_{i+1i+2}\check{R}_{ii+1}=\check{R}_{i+1i+2}\check{R}_{ii+1}\check{R}_{i+1i+2}.
\end{eqnarray}
$\check{R}_{ii+1}$ has the inverse $\check{R}_{ii+1}^{-1}=e_{i}+q^{-1}$ from the \textit{unitary} relation 
$\check{R}_{ii+1}\check{R}_{ii+1}^{-1}=\check{R}_{ii+1}^{-1}\check{R}_{ii+1}=1$. 

Let us introduce the permutation $\mathcal{P}_{ij}$ which exchange the indices $i$ and $j$.  
The $R$-matrix is defined by $R_{ii+1}=\check{R}_{ii+1}\mathcal{P}_{ii+1}$ and obeys the Yang-Baxter equation
\begin{eqnarray}
R_{12}R_{13}R_{23}=R_{23}R_{13}R_{12}. 
\end{eqnarray}

The trigonometric $\check{R}$-matrix can be constructed through the Baxterization of the 
$\check{R}$-matrix:
\begin{eqnarray}\label{def-R}
\check{R}_{i}(u)=\frac{1}{\zeta(u)}\left( u^{1/2}\check{R}_{ii+1}-u^{-1/2}\check{R}_{ii+1}^{-1} \right)
\end{eqnarray}
where $\zeta(u)=qu^{-1/2}-q^{-1}u^{1/2}$. The inverse of $\check{R}_{i}(u)$ is given by 
$\check{R}_{i}^{-1}(u)=\check{R}_{i}(u^{-1})$ and satisfies the \textit{unitary} relation 
$\check{R}_{i}(u)\check{R}_{i}(u^{-1})=1$. 

If we rewrite Eqn.(\ref{def-R}) in terms of the affine Hecke algebra, we have
\begin{eqnarray}\label{def-tri-R}
\check{R}_{i}(z,w):=\check{R}_{i}\left(u=\frac{z}{w}\right)=\frac{qz-q^{-1}w}{qw-q^{-1}z}\mathbf{1}
+\frac{z-w}{qw-q^{-1}z}e_{i}.
\end{eqnarray}
The Yang-Baxter equation for $R_{i}(u)=\check{R}_{i}(u)\mathcal{P}$ is written as
\begin{eqnarray}\label{YB}
\check{R}_{ii+1}\left(\frac{z}{w}\right)\check{R}_{i+1i+2}(z)\check{R}_{ii+1}(w)
=\check{R}_{i+1i+2}(w)\check{R}_{ii+1}(z)\check{R}_{i+1i+2}\left(\frac{z}{w}\right).
\end{eqnarray}

\subsubsection{$q$-Symmetrizers in terms of $\check{R}$}
We will show that the $q$-symmetrizers $Y_{k}$ can be expressed in terms of the 
$\check{R}$-matrix. 
For later convenience, we define 
\begin{eqnarray}\label{L-check}
\check{L}_{i}(m):=\frac{1}{\mu_{m+1}}\check{R}\left(\frac{z}{w}=q^{-2m}\right)=e_{i}-\mu_{m}
\end{eqnarray}
for $m\in\mathbb{N}=\{1,2,\ldots\}$. 

The recursive relation of the $q$-symmetrizer $Y_{m+1}$ (\ref{q-Sym-rr}) is rewritten as 
\begin{eqnarray}\label{q-sym-R}
Y_{m+1}(e_{i},\cdots,e_{i+m})=Y_{m}(e_{i},\cdots,e_{i+m-1})\check{L}_{i+m}(m+1)Y_{m}(e_{i},\cdots,e_{i+m-1}). 
\end{eqnarray}
with $Y_{1}(e_{i})=\check{L}_{i}(1)$. From Eqn.(\ref{q-sym-R}), a $q$-symmetrizer 
is written as a product of $\check{L}_{i}$'s. 

The Hecke relation $e_{i}e_{i\pm1}e_{i}-e_{i}=e_{i\pm1}e_{i}e_{i\pm1}-e_{i\pm1}$ is rewritten in terms of 
the $q$-symmetrizer as $Y_{2}(e_{i},e_{i\pm1})=Y_{2}(e_{i\pm1},e_{i})$. This relation is equivalent 
to the Yang-Baxter equation (\ref{YB}) with a specialization of the spectral parameters. The Yang-Baxter
equation in terms of $\check{L}$-matrix is written as    
\begin{eqnarray}\label{YB2}
\check{L}_{ii+1}(u-v)\check{L}_{i+1i+2}(u)\check{L}_{ii+1}(v)=
\check{L}_{i+1i+2}(v)\check{L}_{ii+1}(u)\check{L}_{i+1i+2}(u-v). 
\end{eqnarray}
where $u,v\in\mathbb{N}$. The Hecke relation is obtained by setting $(u,v)=(2,1)$.

\subsection{Graphical representation of $q$-symmetrizers}\label{subsec-graph-Y}
In this subsection, we introduce the graphical representation of the Yang-Baxter equation (\ref{YB2}). Then, 
we also consider the graphical representation of the $q$-symmetrizers by using the notation used in the above 
subsection. It is well-known that the Yang-Baxter equation for the IRF model (see for example~\cite{Pas88}) can be expressed as the equivalence between different rhombus tilings of a hexagon.
In our novel graphical depiction, the Yang-Baxter equation is also expressed as the equivalence of two 
hexagons. There are, however, nice features in our method. A rhombus represents the $\check{R}$-matrix. 
An integer on the face of the rhombus indicates the spectral parameter of the $\check{R}$-matrix. Furthermore, 
the $q$-symmetrizers are expressed as polygons. These properties play an important role when we construct 
a state of the $A_{k}$ generalized model in Section~\ref{states-Ak}.

\paragraph{The Yang-Baxter equation}
A rhombus represents a $\check{L}$-matrix having the spectral parameter on the face of it:
\begin{eqnarray}\label{L-rhombus}
\check{L}_{i}(m)=\mbox{\raisebox{-0.45\height}{\scalebox{0.6}{\includegraphics{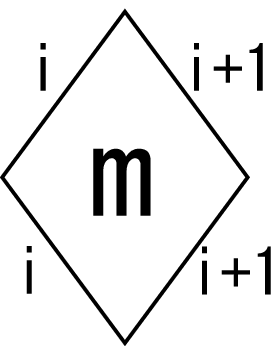}}}}
\end{eqnarray}
We call the edges of a rhombus in Eqn.(\ref{L-rhombus}) the $i$-th and $(i+1)$-th edges. 
The $i$-th and $(i+1)$-th edges indicate the index $i$ of $\check{L}_{i}$. We call the two corners put between 
$i$-th and $(i+1)$-th edges up and down corners, whereas the other two corners right and left corners.
We may omit the name of edges without any confusion. 

Accordingly, the Yang-Baxter equation (\ref{YB2}) of the Hecke type 
is graphically expressed as 
\begin{eqnarray}\label{YB-graph}
\mbox{\raisebox{-0.5\height}{\scalebox{1.2}{\includegraphics{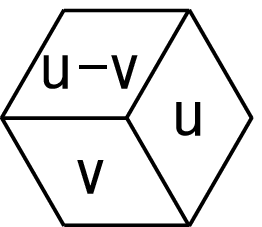}}}}=
\mbox{\raisebox{-0.5\height}{\scalebox{1.2}{\includegraphics{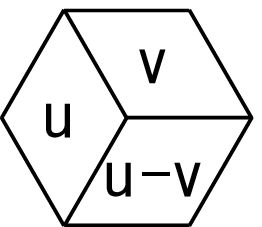}}}}.
\end{eqnarray}
Note that the $\check{L}_{ii+1}$-matrix acts on the $i$-th and $(i+1)$-th edges. 
The order of piling rhombi from the bottom corresponds to the order of $\check{L}$ from right to 
left (since we consider the left ideal later). A rhombus with $u-v$ in l.h.s. of Eqn.(\ref{YB-graph}) is 
piled in the way that the bottom $i$-th and $(i+1)$-th edges of the rhombus are attached to the $i$-th 
and $(i+1)$-th edges of the other two rhombi. 

\paragraph{The $q$-symmetrizer $Y_{k}$} 
As mentioned above, the $q$-symmetrizer $Y_{2}$ is obtained by choosing the special values of the spectral 
parameters in the Yang-Baxter equation. From the fact that each side of the Yang-Baxter equation is graphically 
expressed as a rhombus tiling of a hexagon, $Y_{2}$ is also expressed graphically as the hexagon. 
More generally, we will see that the $q$-symmetrizer $Y_{k}$ has a graphical representation 
by a $2(k+1)$-gon.

Without loss of generality, we consider $Y_{k}=Y_{k}^{(1)}$. We rewrite the $q$-symmetrizer $Y_{k}$ as 
\begin{eqnarray}
Y_{k}&=&Y_{k-1}\check{L}_{k}(k)Y_{k-1} \nonumber \\
&=&\alpha_{k-2}Y_{k-2}\check{L}_{k-1}(k-1)\check{L}_{k}(k)Y_{k-1}\nonumber \\
&=&C \check{L}_{1}(1)\check{L}_{2}(2)\ldots\check{L}_{k}(k)Y_{k-1}
\label{Y-for-graph}
\end{eqnarray}
where $C$ is a constant written in terms of $\alpha_{l}$, $1\le l\le k-1$. 

Eqn.(\ref{Y-for-graph}) implies that $Y_{k}$ is obtained by piling $k$ rhombi corresponding to a sequence 
of $\check{L}$ over a $2k$-gon corresponding to $Y_{k-1}$. Then, we have a $2(k+1)$-gon for the $q$-symmetrizer
$Y_{k}$. 

\bigskip\noindent
{\bf Example:} The $q$-symmetrizers $Y_{3}$ and $Y_{4}$ are expressed as an octagon and a decagon, respectively. 
\begin{figure}[htbp]
\begin{center}
\scalebox{0.5}{\includegraphics{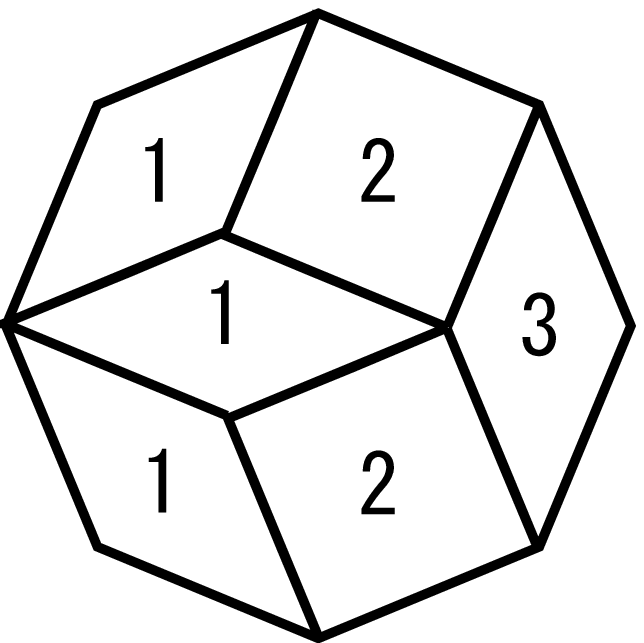}} \ \ \ \ \ \ 
\scalebox{0.4}{\includegraphics{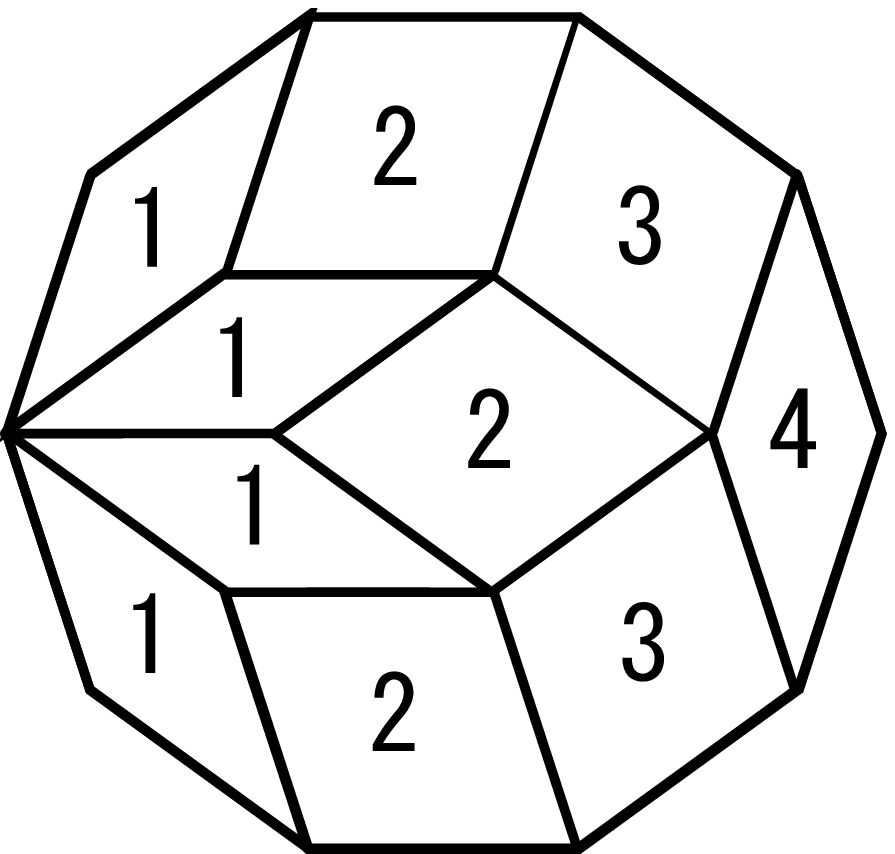}}
\end{center}
\caption{The graphical representation of the $q$-symmetrizers $Y_{3}$ and $Y_{4}$.}
\label{graphical-q-sym-eps}
\end{figure}

\begin{figure}[htbp]
  \begin{center}
   \scalebox{0.5}{\includegraphics{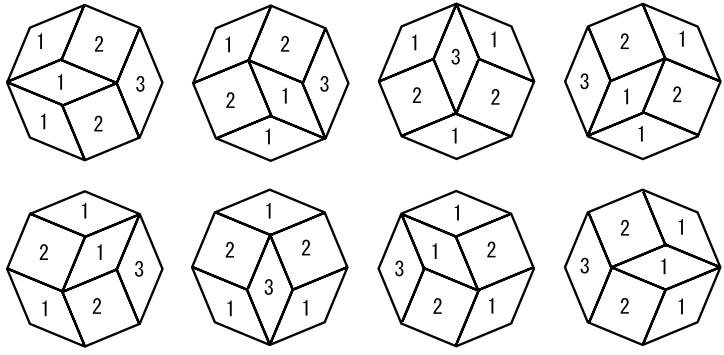}}
  \end{center}
  \caption{Equivalent expressions of the $q$-symmetrizer $Y_{3}$. Elementary moves of rhombus tilings are 
  realized by the Yang-Baxter equation (\ref{YB-graph}).}  \label{fig:q-sym.eps}
\end{figure}

\paragraph{Equivalent expressions of $Y_{k-1}$}
We have seen that the $q$-symmetrizer $Y_{k-1}$ corresponds to a rhombus tiling of a $2k$-gon. However, 
we have many other ways of equivalent rhombus tiling of the $2k$-gon under a sequence of 
\textit{elementary moves} of rhombi. There are two equivalent ways of rhombus tiling 
of a hexagon as in Eqn.(\ref{YB-graph}). An elementary move is an operation which changes a way of 
tiling from l.h.s to r.h.s and vice versa in Eqn.(\ref{YB-graph}). Figure~\ref{fig:q-sym.eps} shows 
all equivalent rhombus tilings of an octagon for the $q$-symmetrizer $Y_{3}$. 

\section{Spin Representation}\label{sec-Spin}
In this section, we will consider the spin representation of the affine Hecke algebra $\widehat{H_{N}^{(k)}}$ with 
the cylindric relation (\ref{cyclic-1}) or (\ref{cyclic-general}). We first introduce the well-known spin 
representation of the Hecke algebra~\cite{Jim86,Mart92}, then introduce the affine generator by the twist. We show that the 
obtained spin representation actually satisfies the defining relations of the Hecke algebras 
$H_{N}^{(k)}(\tau)$ and the cylindric relations. 

\subsection{Hecke algebra}
We first consider the spin representation of the Hecke algebra $H_{N}^{(k)}(\tau)$ and show the 
quotient relation.

Let us consider a representation $(\chi,V^{\otimes N})$ of the quotient Hecke algebra $H_{N}^{(k)}(\tau)$ 
where $\chi: H_{N}^{(k)}(\tau)\rightarrow \mathrm{End}(V^{\otimes N})$ and $V\cong\mathbb{C}^{k}$ is a 
vector space with the standard orthonormal basis $\{|i\rangle \vert 1\le i\le k\}$.  
We denote $|i_{1}\rangle\otimes|i_{2}\rangle\otimes\cdots|i_{n}\rangle\in V^{\otimes n}$ by $|i_{1}i_{2}\cdots i_{n}\rangle$ for brevity. $\langle i|$ is the dual base of $|i\rangle$ with the inner 
product $\langle i|j\rangle=\delta_{ij}$. 

We introduce  $\breve{e}\in\mathrm{End}(V^{\otimes 2})$ which acts on $|ij\rangle$ as
\begin{eqnarray}\label{def-local-e}
\breve{e}|ij\rangle=(1-\delta_{ij})((-q)^{\mathrm{sign}[j-i]}|ij\rangle+|ji\rangle), 
\end{eqnarray}
or equivalently, in terms of the standard basis of $\mathfrak{gl}_{k}$: 
\begin{eqnarray}
\breve{e}=\sum_{a,b=1}^{k}E_{ab}\otimes E_{ba}-\sum_{a,b=1}^{k}q^{\mathrm{sign}[b-a]}E_{aa}\otimes E_{bb}
\end{eqnarray}
where $E_{ab}$ is a $k\times k$ matrix whose elements are $(E_{ab})_{ij}=\delta_{ai}\delta_{bj}$. 

A generator of the Hecke algebra has a representation of $\mathrm{End}(V^{\otimes N})$ 
and is written in terms of $\breve{e}$:
\begin{eqnarray}
\chi(e_{i})=\underbrace{\mathbb{I}\otimes\cdots\otimes\mathbb{I}}_{i-1}\otimes\breve{e}\otimes
\underbrace{\mathbb{I}\otimes\cdots\otimes\mathbb{I}}_{N-i-1}
\end{eqnarray}
where $\mathbb{I}$ is the $k\times k$ identity matrix. Below, we write as $e_{i}$ instead of $\chi(e_{i})$. 

It is straight forward to show that $e_{i}$ satisfies the defining relations of the Hecke 
algebra (\ref{Hecke-relation}). We need to show that this representation actually satisfies the quotient relation. 

\begin{proposition}(\cite{Mart92})\label{vanish-rep}
In the representation $(\chi,V^{\otimes N})$, the generators $e_{i}$'s satisfy the 
quotient relation
\begin{eqnarray}
Y_{k}(e_{i},\cdots,e_{i+k-1})=0, \ \ \mathrm{for}\ i=1,\cdots,N-k.
\end{eqnarray}
\end{proposition}
\begin{proof}
Let us denote $Y_{k}(e_{1},\cdots,e_{k})$ by $Y_{k}^{(1)}$ and $|v\rangle=
|v_{1}\cdots v_{k+1}\rangle\otimes |w\rangle\in V^{\otimes N}$ with $|w\rangle\in V^{\otimes N-k-1}$. 
Since $|w\rangle$ is invariant under the action of $Y_{k}^{(1)}$ from the matrix representation of $e_{i}$'s,  
it is enough to show that $Y_{k}^{(1)}|v\rangle=0$ for any set $\{v_{1},\cdots,v_{k+1}\}$ 
with $1\le v_{i}\le k$. We use the method of induction. Assume that the statement is true up to $k-1$. 
The action of $e_{i}$, $1\le i\le k$, 
on $|v\rangle$ is $e_{i}|v\rangle=0$ if $v_{i}=v_{i+1}$. 
From Prop.~\ref{prop-qY}, we have 
\begin{eqnarray}
Y_{k}^{(1)}|v\rangle=\frac{1}{\alpha_{k-1}}Y_{k}^{(1)}Y_{k-1}^{(1)}|v\rangle
=\frac{1}{\alpha_{k-1}}Y_{k}^{(1)}Y_{k-1}^{(2)}|v\rangle. 
\end{eqnarray}
Since $Y_{k}^{(1)}|v\rangle$ is non-vanishing if $\{v_{1},v_{2},\cdots,v_{k}\}$ and 
$\{v_{2},\cdots,v_{k+1}\}$ are all distinct from the assumption, $v_{1}$ is to be equal to $v_{k+1}$. 
From (\ref{def-local-e}), we have 
\begin{eqnarray*}
Y_{k}^{(1)}|v\rangle&=&\frac{1}{\alpha_{1}}Y_{k}^{(1)}e_{1}|v\rangle \\
&=&\frac{1}{\alpha_{1}}Y_{k}^{(1)}((-q)^{\mathrm{sign}[v_{2}-v_{1}]}|v\rangle+|v_{2}v_{1}v_{3}\cdots v_{k+1}\rangle\otimes|w\rangle \\
&=&\frac{(-q)^{\mathrm{sign}[v_{2}-v_{1}]}}{\alpha_{1}}Y_{k}^{(1)}|v\rangle
\end{eqnarray*}
where we have used $Y_{k-1}^{(2)}|v_{2}v_{1}v_{3}\cdots v_{k+1}\rangle\otimes|w\rangle=0$. Since 
$\frac{(-q)^{\mathrm{sign}[v_{2}-v_{1}]}}{\alpha_{1}}\neq1$ for general $q$ ($q\neq0$), we have 
$Y_{k}^{(1)}|v\rangle=0$. 
\end{proof}

The $q$-symmetrizer satisfies the following properties in the spin representation. 
Below, we restrict the action of $Y_{k-1}^{(i)}$ to $W=V^{\otimes k}$ where 
$V^{\otimes N}=V^{\otimes i-1}\otimes W\otimes V^{\otimes N-i-k}$, since $Y_{k-1}^{(i)}$ acts 
as identity except on $W$. 
\begin{proposition}\label{eigenvec-Y}
For a given $k$, the $q$-symmetrizer $Y_{k-1}^{(i)}$ has only one eigenvector (up to normalization) 
with a non-zero eigenvalue in $V^{\otimes k}$ and its eigenvalue is $\alpha_{k-1}$ given 
by (\ref{eigenvalue-Y}). 
\end{proposition}
\begin{proof}
We will show that we have a simultaneous eigenvector of $e_{i}$'s and that is also the only eigenvector 
of the $q$-symmetrizer. 
From the spin representation, we only need to prove that $Y_{k-1}=Y_{k-1}^{(1)}$ has only one eigenvector with 
a non-zero eigenvalue. Let $v=|v_{1}\cdots v_{k}\rangle$ be a vector in $V^{\otimes k}$. 
From Proposition~\ref{vanish-rep}, a non-vanishing $v$ satisfies $v_{i}\neq v_{j}$ for any $i,j$. 
We denote by $\widetilde{V^{\otimes k}}$ the subspace of $V^{\otimes k}$ where $1,\cdots,k$ appear exactly once 
in $\{v_{1},\cdots,v_{k}\}$. Note that non-vanishing eigenvectors of $Y_{k-1}$ are in 
$\widetilde{V^{\otimes k}}$. From Proposition~\ref{prop-qY}, eigenvectors of $Y_{k-1}$ are also 
simultaneous eigenvectors of all the Hecke generators $e_{i}$'s and vice versa. 
The eigenvectors of $e_{i}$ with non-zero eigenvalue have the form 
\begin{eqnarray}
|\cdots v_{i}v_{i+1}\cdots\rangle+(-q)^{\mathrm{sign}[v_{i}-v_{i+1}]}|\cdots v_{i+1}v_{i}\cdots\rangle.
\end{eqnarray}
Starting from $|v\rangle=|v_{1}\cdots v_{k}\rangle=|12\ldots k\rangle\in\widetilde{V^{\otimes k}}$, 
we may fix the simultaneous eigenvector of all $e_{i}$'s (up to the overall constant) as follows:
\begin{eqnarray}\label{spin-eigenvec-Y}
|v_{0}\rangle=\sum_{s\in\mathfrak{S}_{k}}(-q)^{l(s)}|s(v)\rangle, 
\end{eqnarray}
where $v_{0}\in\widetilde{V^{\otimes k}}$, $\mathfrak{S}_{k}$ is the symmetric group and 
$|s(v)\rangle=|v_{s(1)}v_{s(2)}\cdots v_{s(k)}\rangle$. The function $l(s)$ satisfies 
$l(ss^{'})=l(s)+l(s^{'})$ and 
\begin{eqnarray}\label{def-ls}
l(s_{ii+1})=\left\{
  \begin{array}{cc}
    1,   & v_{i}>v_{i+1}   \\
    -1,   &  v_{i+1}>v_{i}  \\
  \end{array}
\right.,
\end{eqnarray}
where $s_{ii+1}\in\mathfrak{S}_{k}$ is the transposition between $v_{i}$ and $v_{i+1}$. 
Another choice of $|v\rangle$ gives just the difference of overall normalization constant. 
Since $v_{0}$ is constructed as the simultaneous eigenvector of $e_{i}$'s, 
$v_{0}$ is also the eigenvector of $Y_{k}$. The uniqueness of the eigenvector is guaranteed by construction. 

The eigenvalue of $|v_{0}\rangle$ with respect to $e_{i}$ is $\tau$. Then, the action of $Y_{k-1}$ on 
$|v_{0}\rangle$ is $Y_{k-1}|v_{0}\rangle=\tilde{y}_{k-1}|v_{0}\rangle$ where 
$\tilde{y}_{k-1}=Y_{k-1}(\tau,\cdots,\tau)$ is a $c$-number. $\tilde{y}_{k-1}$ satisfies the same 
recurrence relation and the initial condition as (\ref{recur-alpha}). This means $\tilde{y}_{k-1}=\alpha_{k-1}$. 
\end{proof}

For later convenience, we write down the inner product of $|v_{0}\rangle$ (the eigenvector of 
$Y_{k-1}$ with non-zero eigenvalue) in terms of the Chebyshev polynomials of the second kind.
\begin{corollary}\label{inner-eigenvec}
Let $|v_{0}\rangle$ be the eigenvector of the $q$-symmetrizer $Y_{k}$ with the non-zero eigenvalue. 
The inner product $I_{k}=\langle v_{0}|v_{0}\rangle$ in the representation $(\chi,V^{\otimes N})$ 
is calculated as $q^{-k(k-1)/2}\prod_{i=1}^{k}U_{i}$. 
\end{corollary}
\begin{proof}
From Proposition~\ref{eigenvec-Y}, the eigenvector $|v_{0}\rangle$ can be rewritten as
\begin{eqnarray}
|v_{0}\rangle&=&\sum_{s\in\mathfrak{S}_{k}}(-q)^{l(s)}|s(v)\rangle \\
&=&\sum_{1\le v_{1}\le k}\sum_{\tilde{s}\in\mathfrak{S}_{k-1}}(-q)^{-(v_{1}-1)+l(\tilde{s})}
|v_{1}\tilde{s}(v\setminus v_{1})\rangle
\end{eqnarray}
where $v=|12\ldots k\rangle$ and 
$|v_{1}\tilde{s}(v\setminus v_{1})\rangle=|v_{1}v_{\tilde{s}(2)}\cdots v_{\tilde{s}(k)}\rangle$. The inner 
product  $I_{k}$ is then calculated as
\begin{eqnarray}
I_{k}&=&\langle v_{0}|v_{0}\rangle =\sum_{s\in\mathfrak{S}_{k}}(-q)^{2l(s)} \nonumber \\
&=&\sum_{1\le v_{1}\le k}q^{-2(v_{1}-1)}\sum_{\tilde{s}\in\mathfrak{S}_{k-1}}q^{2l(\tilde{s})} 
=(-q)^{-(k-1)}U_{k-1}I_{k-1} \nonumber \\
&=&(-q)^{-k(k-1)/2}\prod_{1\le i\le k-1}U_{i}.
\end{eqnarray}
\end{proof}

\subsection{Affine Hecke algebra}
Let us introduce a linear operator $\tilde{e}\in\mathrm{End}(V\otimes V)$ in the basis of 
$\mathfrak{gl}_{k}$ by
\begin{eqnarray}\label{aff-e}
\tilde{e}=\sum_{a,b=1}^{k}q^{2(a-b)}E_{ab}\otimes E_{ba}
-\sum_{a,b=1}^{k}q^{\mathrm{sign}[b-a]}E_{aa}\otimes E_{bb}.
\end{eqnarray}
This $\tilde{e}$ is obtained by the twist, {\it i.e.}, 
$\tilde{e}=\Omega^{-1}\breve{e}\Omega$ where the twist $\Omega$ 
is given by $\Omega=\mathbb{I}\otimes\tilde{\Omega}$, 
$\tilde{\Omega}=\mbox{diag}(q^{-(k-1)},q^{-(k-3)},\cdots,q^{k-1})$.

It is also straightforward to show that the spin representation (\ref{aff-e}) of $\tilde{e}$ satisfies 
the following two properties aff\ref{aff1}-\ref{aff2}. 
\begin{enumerate}[({aff}1)]
  \item \label{aff1}$\tilde{e}^{2}=\tau \tilde{e}$
  \item \label{aff2} $\tilde{e}_{12}e_{23}\tilde{e}_{12}-\tilde{e}_{12}=e_{12}\tilde{e}_{23}e_{12}-e_{12}$
\end{enumerate}
where $\tilde{e}$ is a $k\times k$ matrix and $\tilde{e}_{12}=\tilde{e}\otimes\mathbb{I}$, 
$\tilde{e}_{23}=\mathbb{I}\otimes\tilde{e}$ and $\tilde{e}_{13}=\mathcal{P}_{23}\tilde{e}_{12}\mathcal{P}_{23}$ 
($\mathcal{P}$ is a permutation matrix).

Now we are ready to construct an additional generator $e_{N}$ which allows us to have 
the affine Hecke algebra. Let us introduce the shift operator $\rho$ acting on the basis 
in $V^{\otimes N}$. Let $v=|v_{1}\cdots v_{N}\rangle$ be a base in $V^{\otimes N}$. Then, 
$\rho: V^{\otimes N}\rightarrow V^{\otimes N}$ is defined by $\rho:v\mapsto |v_{2}\cdots v_{N}v_{1}\rangle$. 
We define $e_{N}$ acting on $V^{\otimes N}$ as 
\begin{eqnarray}
e_{N}=\rho^{-1}(\underbrace{\mathbb{I}\otimes\cdots\otimes\mathbb{I}}_{N-2}\otimes\tilde{e})\rho.
\end{eqnarray}

By construction, the defining relations (\ref{Hecke-relation}) of the Hecke algebra become cyclic and there exists 
the cyclic operator $\sigma$ such that $\sigma e_{i}=e_{i-1}\sigma$ for any $i\in \mathbb{Z}/N\mathbb{Z}$. 
The cyclic operator $\sigma$ in the spin representation is explicitly given by 
\begin{eqnarray}\label{sigma-spinrep}
\sigma=(\Omega\rho)^{-1}.
\end{eqnarray}
Here $\Omega=\mathbb{I}^{\otimes N-2}\otimes\tilde{\Omega}$. Note that $\sigma^{N}$ is the identity.  
In this way, we construct the affine Hecke generators $\{e_{1},\cdots,e_{N},\sigma\}$ in the spin representation.

\bigskip
\noindent
{\bf Remark:}
There may be other linear operators in $\mathrm{End}(V\otimes V)$ which satisfy the properties aff\ref{aff1} and
aff\ref{aff2}. If we set $\sigma=\rho^{-1}$ instead of (\ref{sigma-spinrep}), we also have another affine 
Hecke algebra. However, this algebra does not satisfy the cylindric relation (may satisfy another kind of 
vanishing condition). 

\bigskip
The affine algebra $\widehat{H^{(k)}_{N}}$ can be regarded as a natural generalization of the affine TL algebra 
on a cylinder. As we will see in the next paragraph, the affine Hecke algebra $\widehat{H^{(k)}_{N}}$ satisfies 
the cylindric relation introduced in the Section~\ref{sec-Hecke}. 
To relate this affine Hecke algebra to the loop models, we need to have a 
further condition for $\tilde{e}$, which comes from the weight of a ``loop" surrounding a cylinder. 
Although the graphical way to describe a ``loop" model corresponding to the $A_{k}$-vertex model is not known as 
far as the authors know, it is natural to assign that the weight of a ``loop" is related to the Chebyshev 
polynomials of the second kind. This is realized by the cylindric relation (see Section~\ref{subsec-Akmodel}). 

Let us consider the case $k=2$. The spin representation of $\tilde{e}$ is given by
\begin{eqnarray}
\tilde{e}= \left(
  \begin{array}{cccc}
 0 & 0 & 0 & 0 \\
 0 &  -q  & q^{-2} & 0  \\
 0 &  q^{2} & -q^{-1} & 0  \\
 0 & 0 & 0 & 0
  \end{array}
\right), 
\end{eqnarray}
Note that $\tilde{e}=\Omega^{-1}\breve{e}\Omega$ where $\Omega=\mathbb{I}\otimes \tilde{\Omega}$ and 
$\tilde{\Omega}=\mbox{diag}(q^{-1},q)$, that is, $\tilde{e}$ is the twist of $\breve{e}$. This 
$\tilde{e}$ is equivalent to the boundary considered in~\cite{MitNiedeGieBat04,DiFZJZu06}. 

\paragraph{Cylindric relation and extra vanishing conditions}
We need to show that the above spin representation satisfies the cyclicity of the vanishing conditions (\ref{vc-Hecke}) 
and the cylindric relation (\ref{cyclic-1}) or (\ref{cyclic-general}). 

The vanishing conditions (\ref{vc-Hecke}) of the quotient Hecke algebra become cyclic by adding 
the extra generator $e_{N}$ when $n\ge 2$. We omit the proof but similar to the Proposition~\ref{vanish-rep} 
because of the properties aff\ref{aff1} and aff\ref{aff2}.

We will show that the spin representation satisfies the cylindric relations (\ref{cyclic-1}) or 
(\ref{cyclic-general}) in the following two propositions. 
\begin{proposition}
In the representation $(\chi,V^{\otimes N})$ of the affine Hecke algebra, the cylindric relation 
for $N=k$ is given by
\begin{eqnarray}\label{cyclic-rel-N-k}
Y_{k-1}(e_{1},\cdots,e_{k-1})(e_{k}-\tau)Y_{k-1}(e_{1},\cdots,e_{k-1})=0.
\end{eqnarray}
\end{proposition}
\begin{proof}
By using Proposition~\ref{eigenvec-Y} and \ref{inner-eigenvec}, the $q$-symmetrizer $Y_{k-1}^{(1)}$ is 
written as 
\begin{eqnarray}\label{relation-Y-v}
Y_{k-1}=\frac{\alpha_{k-1}}{A_{k-1}}|v_{0}\rangle\langle v_{0}|
\end{eqnarray}
where $\alpha_{k-1}$ is given by (\ref{eigenvalue-Y}) and $A_{k-1}=(-q)^{-k(k-1)/2}\prod_{1\le i\le k-1}U_{i}$.  

Equation~(\ref{cyclic-rel-N-k}) is rewritten as 
$\langle v_{0}|e_{k}|v_{0}\rangle=\tau \langle v_{0}|e_{k}|v_{0}\rangle$. 
By the definition $|v_{0}\rangle=\sum_{s\in\mathfrak{S}_{k}} (-q)^{l(s)}|s(v)\rangle$, we have 
\begin{eqnarray}\label{vev-ek}
\langle v_{0}|e_{N}|v_{0}\rangle=\sum_{s,s'\in\mathfrak{S}_{k}}(-q)^{l(s)+l(s')}\langle s'(v)|e_{k}|
s(v)\rangle .
\end{eqnarray}
Due to the representation (\ref{aff-e}) of $e_{k}$, the expectation value $\langle s'(v)|e_{k}|s(v)\rangle$ 
is shown to be non-zero for either $s'=s$ or $s'=s_{1k}s$ where $s_{1k}$ is the transposition operator.  
Therefore, we have 
\begin{eqnarray}\label{l-ldash}
l(s')=\left\{
  \begin{array}{cc}
    l(s),   &  s'=s  \\
    l(s)-2(v_{s(k)}-v_{s(1)})+\mathrm{sign}[v_{s(k)}-v_{s(1)}],   &  s'=s_{1k}s  \\
  \end{array}
\right.
\end{eqnarray}
and 
\begin{eqnarray}\label{vev-ek2}
\frac{\langle s'(v)|e_{k}|s(v)\rangle}{\langle s'(v)|s(v)\rangle}=\left\{
  \begin{array}{cc}
    (-q)^{\mathrm{sign}[v_{s(1)}-v_{s(k)}]},   &  s'=s   \\
    (-q)^{2(v_{s(k)}-v_{s(1)})},   & s'=s_{1k}s   \\
  \end{array}
\right.
\end{eqnarray}
Substituting Eqn.(\ref{vev-ek2}) and Eqn.(\ref{l-ldash}) into Eqn.(\ref{vev-ek}), we finally obtain
\begin{eqnarray}
\langle v_{0}|e_{k}|v_{0}\rangle&=&
\sum_{s,s'\in\mathfrak{S}_{k}}(-q)^{l(s)+l(s')}\langle s'(v)|e_{k}|s(v)\rangle \nonumber \\
&=&\sum_{s\in\mathfrak{S}_{k}}(-q)^{2l(s)}(-q-q^{-1})\langle s(v)|s(v)\rangle \nonumber \\
&=&\tau \langle v_{0}|v_{0}\rangle. 
\end{eqnarray}
\end{proof}

\begin{proposition}
In the case of $N=nk$, $n\ge 2$, the affine Hecke generators $\{e_{1},\cdots,e_{N}\}$ in the 
representation $(\chi,V^{\otimes N})$ satisfy the following relation:
\begin{eqnarray}\label{cyclic-rel-N}
Y_{q\text{\rm -sym}}\cdot \prod_{i=1}^{n-1}(e_{ik}-\mu_{k-1})\cdot(e_{nk}-\tau)\cdot Y_{q\text{\rm -sym}}=0,
\end{eqnarray}
where $Y_{q\text{\rm -sym}}=\prod_{i=0}^{n-1}Y_{k-1}^{(ik+1)}$. 
\end{proposition}
\begin{proof}
We first rewrite the relation (\ref{cyclic-rel-N}) into a simpler form.

Let $|v_{0}^{(i+1)}\rangle=\sum_{s\in\mathfrak{S}_{k}}(-q)^{l(s)}|v_{s(1)}^{(i)}\cdots v_{s(k)}^{(i)}\rangle\in V^{\otimes k}$ 
be the eigenvector of $Y_{k-1}^{(ik+1)}$ for $0\le i\le n-1$ (see Proposition~\ref{eigenvec-Y}). 
The unique eigenvector $|v_{0}\rangle\in V^{\otimes N}$ of $Y_{q\text{-sym}}$ is given by the tensor product 
of $|v_{0}^{(i)}\rangle$, {\textit i.e.} $|v_{0}\rangle=\bigotimes_{i=0}^{n-1}|v_{0}^{(i+1)}\rangle$. By taking 
the expectation value w.r.t. $|v_{0}\rangle$, Eqn.(\ref{cyclic-rel-N}) is rewritten 
as
\begin{eqnarray}\label{cyclic-vev}
\Big\langle \prod_{i=1}^{n-1}(e_{ik}-\mu_{k-1})\cdot(e_{nk}-\tau)\Big\rangle=0, 
\end{eqnarray}
where we denote $\langle v_{0}|{\cal{O}}|v_{0}\rangle$ for some operator ${\cal{O}}$ by $\langle{\cal{O}}\rangle$.  

The vanishing condition $Y_{k}^{(ik+1)}=0$ for any $i\in\mathbb{Z}/k\mathbb{Z}$ is expressed as  
$\langle e_{(i+1)k}-\mu_{k}\rangle=0$. In general, we have 
\begin{eqnarray}\label{vc-prode}
\Big\langle \prod_{j=1}^{m}(e_{i_{m}k}-\mu_{k})\Big\rangle=0
\end{eqnarray}
for $1\le i_{m}\le k$ and $1\le m\le n-1$ where we have used the relation $e_{lk}Y_{k-1}^{(ik+1)}=Y_{k-1}^{(ik+1)}e_{lk}$ for 
$l\neq ik,(i+1)k$. Equation~(\ref{cyclic-vev}) can be rewritten as
\begin{eqnarray}
\mathrm{l.h.s.\ in\ } (\ref{cyclic-vev})&=&\Big\langle\prod_{i=1}^{n}(e_{ik}-\mu_{k})
+\Delta_{k}^{n-1}(\mu_{k}-\tau)\Big\rangle \\
&=&\Big\langle \prod_{i=1}^{n}e_{ik}-\mu_{k}^{n}+\Delta_{k}^{n-1}(\mu_{k}-\tau)\Big\rangle
\end{eqnarray}
where we have used Eqn.(\ref{vc-prode}), $\langle\prod_{i=1}^{m} e_{ik}\rangle=\mu_{k}^{m}$, $1\le m\le k-1$,  
and $\Delta_{k}=\mu_{k}-\mu_{k-1}$. By the relations $\tau-\mu_{k}=\mu_{k+1}^{-1}$ 
and $U_{k-1}^{2}-U_{k}U_{k-2}=1$, eventually the wanted relation equivalent to Eqn.(\ref{cyclic-rel-N}) is equivalent to
\begin{eqnarray}
\Big\langle\prod_{i=1}^{n}e_{ik}\Big\rangle=\frac{1}{U_{k}^{n}U_{k-1}^{n}}
(U_{k-1}^{2n}+U_{k-1}U_{k+1})\langle v_{0}|v_{0}\rangle.
\end{eqnarray}
\medskip

On the other hand, we can evaluate $\langle\prod_{i=1}^{n}e_{ik}\rangle$ by using the spin representation. Let us 
consider the action of $\prod_{i=1}^{n}e_{ik}$ on the vector 
$|v\rangle=|v_{1}^{(1)}\ldots v_{k}^{(1)}v_{1}^{(2)}\ldots v_{k}^{(n)}\rangle
\in\left(\widetilde{V^{\otimes k}}\right)^{\otimes n}$. 
The operator $e_{ik}$ acts locally on $v_{k}^{(i)}$ and $v_{1}^{(i+1)}$. 
To have a non-vanishing expected value, $\langle v'|\prod_{i=1}^{k}e_{ik}|v\rangle\neq0$, an admissible 
$\langle v'|$ satisfies ${v'}_{m}^{(i)}=v_{m}^{(i)}$ for all $i\in\mathbb{Z}/n\mathbb{Z}$, $2\le m\le k-1$ and 
either of the following conditions:
\begin{itemize}
\item ${v'}^{(i)}_{1}=v^{(i)}_{1}$ and ${v'}^{(i)}_{k}=v^{(i)}_{k}$ for all $i\in\mathbb{Z}/n\mathbb{Z}$,
\item ${v'}^{(i)}_{k}=v^{(i+1)}_{1}$ and ${v'}^{(i+1)}_{1}=v^{(i)}_{k}$ for all $i\in\mathbb{Z}/n\mathbb{Z}$.
\end{itemize}

Now, we are ready to evaluate 
\begin{eqnarray}\label{vev-prod-e-2}
\left\langle\prod_{i=1}^{n}e_{ik}\right\rangle=\sum_{S}(-q)^{\sum_{i}l(s_{i})+l(t_{i})}
\Big\langle\bigotimes_{i=1}^{n}t_{i}(v^{(i)})\Big|\prod_{i=1}^{n}e_{ik}\Big|\bigotimes_{i=1}^{n}s_{i}(v^{(i)})
\Big\rangle. 
\end{eqnarray}
where the sum $S$ is take all over $s_{1},\cdots,s_{n}\in\mathfrak{S}_{k}$ and 
$t_{1},\cdots,t_{n}\in\mathfrak{S}_{k}$. From the above considerations, we split the calculation into two cases 
as follows. 
\bigskip

\underline{Case 1:} We consider the case where $s_{i}(1)=s_{j}(1), s_{i}(k)=s_{j}(k)$ for all $i,j\in\mathbb{Z}/n\mathbb{Z}$ 
and  $t_{i}$'s satisfy
\begin{itemize}
  \item $t_{i}(1)=s_{i-1}(k)$ and $t_{i}(k)=s_{i+1}(1)$ for all $i\in\mathbb{Z}/n\mathbb{Z}$,
  \item $t_{i}(j)=s_{i}(j)$ for $2\le j\le k-1$ and all $i\in\mathbb{Z}/n\mathbb{Z}$.
\end{itemize}
We abbreviate as $t=t_{i}$, $s=s_{i}$ without confusion. 
From (\ref{def-ls}), we have 
\begin{eqnarray}\label{rel-t-s}
l(t)=l(s)-2(v_{s(k)}-v_{s(1)})+\mathrm{sign}[v_{s(k)}-v_{s(1)}].
\end{eqnarray}
If we rewrite $|s(v)\rangle$ in terms of $\tilde{s}\in\mathfrak{S}_{k-2}$ such that 
$|s(v)\rangle\propto |v_{s(1)}\tilde{s}(v)v_{s(k)}\rangle$, we have 
\begin{eqnarray}\label{recur-s-s}
l(s)=l(\tilde{s})+(v_{s(k)}-v_{s(1)})-(k-2)-\frac{1}{2}\mathrm{sign}[v_{s(k)}-v_{s(1)}]-\frac{1}{2}. 
\end{eqnarray}
The action of $\prod_{i=1}^{n-1}e_{ik}$ gives a factor $1$ and that of $e_{nk}$ gives $q^{a}$ where 
\begin{eqnarray}\label{factor-by-e}
a=-2(v_{s(1)}^{(1)}-v_{s(k)}^{(n)}).
\end{eqnarray}
Substituting Eqn.(\ref{rel-t-s}), (\ref{recur-s-s}) and (\ref{factor-by-e}) into (\ref{vev-prod-e-2}), 
we have 
\begin{eqnarray}
\begin{split}
J_{1}&=\sum_{1\le v_{s(k)}^{(n)}\neq v_{s(1)}^{(1)}\le k}\sum_{\tilde{s}_{i}\in\mathfrak{S}_{n-2}\atop 1\le i\le n}
(-q)^{-n(2k-3)+\sum_{i}2l(\tilde{s}_{i})-2(v_{s(1)}^{(1)}-v_{s(k)}^{(n)})} \\
&=(-q)^{-n(2k-3)}I_{k-2}^{n}(U_{k+1}U_{k-1}-(k-1)) \\
&=\frac{1}{U_{k}^{n}U_{k-1}^{n}}(U_{k+1}U_{k-1}-(k-1))\langle v_{0}|v_{0}\rangle
\end{split}\label{j1}
\end{eqnarray}
where we have used the recurrence relation for $I_{k}$ obtained in Proposition~\ref{inner-eigenvec}. 
\bigskip

\underline{Case 2:} We consider the case where $s_{i}(j)=t_{i}(j)$ for all $1\le j\le k$ and $i\in\mathbb{Z}_{n}$. 
Obviously, we have  
\begin{eqnarray}\label{rel-ts-2}
l(t_{i})=l(s_{i})
\end{eqnarray}
 for all $i$. The action of $e_{ik}$'s gives a factor $(-q)^{b}$ where  
\begin{eqnarray}\label{factor-by-e-2}
b=\sum_{i=1}^{n}\mathrm{sign}[v_{s(1)}^{(i+1)}-v_{s(k)}^{(i)}]
\end{eqnarray}
Substituting (\ref{recur-s-s}), (\ref{rel-ts-2}) and (\ref{factor-by-e-2}) into (\ref{vev-prod-e-2}), we have
\begin{align}
\begin{split}
J_{2}&=\sum_{s_{i}\in\mathfrak{S}_{n}}(-q)^{\sum_{i}2l(s_{i})+\mathrm{sign}[v_{s(1)}^{(i+1)}-v_{s(k)}^{(i)}]} \\
&=\sum_{S^{'}} \sum_{\tilde{s}_{i}\in\mathfrak{S}_{k-2}}(-q)^{\sum_{i}2l(\tilde{s}_{i})+2(v_{s(k)}^{(i)}-v_{s(1)}^{(i)})
-2(k-2)-\mathrm{sign}[v_{s(k)}^{(i)}-v_{s(1)}^{(i)}]-1+\mathrm{sign}[v_{s(1)}^{(i+1)}-v_{s(k)}^{(i)}]} \\
&=(-q)^{-n(2k-3)}I_{k-2}\sum_{S^{'}}(-q)^{\sum_{i}2(v_{s(k)}^{(i)}-v_{s(1)}^{(i)})
-\mathrm{sign}[v_{s(k)}^{(i)}-v_{s(1)}^{(i)}]-\mathrm{sign}[v_{s(k)}^{(i)}-v_{s(1)}^{(i+1)}]}
\end{split}\label{j2}
\end{align}
and the sum is taken over all the sets:  
\begin{eqnarray}
S^{'}=\left\{ v_{s(1)}^{(i)},v_{s(k)}^{(i)} \left|
  \begin{array}{c}
    1\le v_{s(1)}^{(i)},v_{s(k)}^{(i)}\le k,\quad i=1,2,\ldots,n\\
    v_{s(1)}^{(i)}\neq v_{s(k)}^{(i)}, v_{s(1)}^{(i)}\neq v_{s(k)}^{(i-1)}  \\
  \end{array}
\right.\right\}. 
\end{eqnarray}
From Lemma~\ref{lemma-for-cyclic} (see below), $J_{2}$ in Eqn.(\ref{j2}) is rewritten as 
\begin{eqnarray}
J_{2}=\frac{1}{U_{k}^{n}U_{k-1}^{n}}(U_{k-1}^{2n}+(k-1)).
\end{eqnarray}

Together with Eqn.(\ref{j1}) and Eqn.(\ref{vev-prod-e-2}), we finally obtain  
\begin{eqnarray}
\left\langle\prod_{i=1}^{n}e_{ik}\right\rangle&=&J_{1}+J_{2} \nonumber \\
&=&\frac{1}{U_{k}^{n}U_{k-1}^{n}}(U_{k-1}^{2n}+U_{k-1}U_{k+1})I_{k}
\end{eqnarray}
and this completes the proof of the relation (\ref{cyclic-rel-N}). 
\end{proof}

We need the following lemma.
\begin{lemma}\label{lemma-for-cyclic}
Let
\begin{eqnarray}
I&=&\sum_{S'}q^{\sum_{l=1}^{n}2(i_{2l}-i_{2l-1})-\mathrm{sign}[i_{2l}-i_{2l-1}]-
\mathrm{sign}[i_{2l}-i_{2l+1}]}
\end{eqnarray}
where $i_{2n+1}=i_{1}$ and 
\begin{eqnarray}
S'=\left\{ i_{l}  \left|
\begin{array}{c}
 1\le i_{l}\le k, \\
 i_{l}\neq i_{l\pm1},  \\ 
\end{array}\  \mathrm{for}\ 1\le \forall l \le 2n  
\right.\right\}. 
\end{eqnarray}
Then, $I$ is calculated in terms of $U'_{k-1}=U_{k-1}(-\tau)$ as 
\begin{eqnarray}
I=(U'_{k-1})^{2n}+(k-1).
\end{eqnarray}
\end{lemma}
\begin{proof}
Let us introduce a set of integer variables, 
$U=\{(u_{1},\ldots,u_{2n})| 1\le u_{j}\le k-1, \mathrm{for}\ 1\le j\le2n\}$ and the subset of $S'$, 
$S'_{extra}=\{ (i_{1}\ldots,i_{2n}) | i_{2l-1}=k^{'}+1, i_{2l}=k^{'}, 1\le k^{'}\le k-1, 
\mathrm{for}\ 1\le l\le n  \}$.  

We introduce the shift operator acting on a sequence of length $2n$, $S=(s_{1},\cdots,s_{2n})$, by 
$\xi:S\rightarrow S, s_{i}\mapsto s_{i+1}$ for $i\in\mathbb{Z}_{2n}$.  We 
consider two subsets $S_{0}\subset S'\backslash S'_{extra}$ and $U_{0}\subset U$:
\begin{eqnarray}
S_{0}&=&\left\{(i_{1},\cdots,i_{2n})\in S'\backslash S'_{extra}  \left| 
\xi^{m}(i_{j})\succeq (i_{j}), \ \ 1\le\forall m\le 2n-1
\right.\right\}, \\
U_{0}&=&\left\{(u_{1},\cdots,u_{2n})\in U  \left| 
\xi^{m}(u_{j})\succeq (u_{j}), \ \ 1\le\forall m\le 2n-1
\right.\right\}. 
\end{eqnarray}
The symbol $\succeq$ means lexicographic order, {\textit i.e.} $\mu\succeq \nu$ stands for 
$\mu_{j}=\nu_{j}$ for all $1\le j\le 2n$, or $\mu_{j}=\nu_{j}$ for $1\le j\le i$ and $\mu_{i+1}>\nu_{i+1}$ for 
some $i$. 
When we have a bijection $\eta:S_{0}\rightarrow U_{0}$, We extend a bijection from 
$S'\backslash S'_{extra}$ to $U$. For a given $i\in S'\backslash S'_{extra}$, we have 
a non-negative integer $r_{min}=\min\{r: \xi^{r}i\in S_{0}\}$. Then a bijection is extended by
$\xi^{-r_{min}}\circ\eta\circ\xi^{r_{min}}$. 

We construct a bijection $\eta:S_{0}\rightarrow U_{0}$ by first constructing an 
injection $\eta:S_{0}\rightarrow U_{0}$ and showing there exists the injective inverse 
$\eta^{-1}$. 

The map $\eta:S_{0}\rightarrow U_{0}$ defines $u_{j}$ recursively starting from $u_{1}$:
\begin{eqnarray}
u_{1}&=&i_{1},\qquad u_{2}=u_{1}+d_{1}', \nonumber \\ 
u_{j}&=&u_{j-1}+
\left\{\begin{array}{cc}
d_{j}, & d_{j}d_{j-1}>0  \\
d_{j}', & d_{j}d_{j-1}<0 \label{branch-d}
\end{array}\right.,  
\end{eqnarray}
where $d_{j}:=i_{j+1}-i_{j}$ and $d_{j}':=i_{j+1}-i_{j}-\mathrm{sign}[i_{j+1}-i_{j}]$ for $1\le j\le 2n-1$. 
From the construction, $\eta$ is injective.  
We have $\mbox{Im}(\eta)\subseteq U_{0}$ since the branching rule (\ref{branch-d}) assures 
$u_{j}\ge u_{1}$ and $\max\{u\}\le \max\{i\}-1\le k-1$. Then, the inverse $\eta^{-1}:U_{0}\rightarrow S_{0}$ is explicitly given by 
\begin{eqnarray}
i_{1}&=&u_{1}, \\
i_{j}&=&i_{j-1}+t_{j-1}
\end{eqnarray}
and 
\begin{eqnarray}
t_{j}=\left\{
\begin{array}{cc}
-\mathrm{sign}[t_{j-1}], & \bar{d}_{j-1}=\bar{d}_{j}=0 \\
-\mathrm{sign}[\bar{d}_{j-1}], & \bar{d}_{j-1}\neq0, \bar{d}_{j}=0 \\
\bar{d}_{j}+\mathrm{sign}[\bar{d}_{j}], & t_{j-1}\bar{d}_{j}>0  \\
\bar{d}_{j}, & t_{j-1}\bar{d}_{j}<0
\end{array}\right.
\end{eqnarray}
where $\bar{d}_{j}=u_{j+1}-u_{j}$  with the initial condition $t_{1}=\bar{d}_{1}+1$.  
The map $\eta^{-1}$ is also injective. It is easy to verify that $\mbox{Im}(\eta^{-1})\subseteq S_{0}$ 
since $i_{2n}\ge i_{1}+1$, $i_{j}\neq i_{j+1}$ for $1\le \forall j\le2n$ and $\max\{i_{j}\}=\max\{u\}+1\le k$. 
From these, $\eta:S_{0}\rightarrow U_{0}$ is a bijection.  

Note that when $i\in S_{0}$ and $u=\eta(i)\in U_{0}$, we have  
\begin{eqnarray}
\sum_{l=1}^{n}2(u_{2l}-u_{2l-1})&=&\sum_{l=1}^{n}(u_{2l}-u_{2l-1})+(u_{2l}-u_{2l+1}) \nonumber \\
&=&\sum_{i=1}^{n}2(i_{2l}-i_{2l-1})-\mathrm{sign}[i_{2l}-i_{2l-1}]-
\mathrm{sign}[i_{2l}-i_{2l+1}], \nonumber
\end{eqnarray}
since the branching rules (\ref{branch-d}) give the correct term for $u_{2l}-u_{1}=i_{2l}-i_{1}-1$. 

From these observations, we arrive at  
\begin{eqnarray}
I&=&\sum_{U}q^{\sum_{l=1}^{2n}(u_{2l}-u_{2l-1})}+\sum_{S_{extra}^{'}}1 \nonumber \\
&=&(U'_{k-1})^{2n}+(k-1). \label{qpoweriu}
\end{eqnarray}
\end{proof}

\paragraph{Example:}The following list gives some examples of the bijection. 
\begin{center}
  \begin{tabular}{c|c|c}
    $i$   &  $u$  & Exponent in Eqn.(\ref{qpoweriu})  \\
    \hline
    $(1, 2, 3, 7, 4, 2)$ & $(1, 1, 2, 6, 4, 2)$ & 4 \\
   $(7, 2, 3, 4, 2, 1)$ &  $(6, 2, 2, 3, 2, 1)$ &  -8 \\
 $(2, 1, 5, 3, 1, 4, 2, 5)$      &  $(2, 1, 4, 3, 1, 3, 2, 4)$  &    4  \\
  \end{tabular}
\end{center}

\noindent
{\bf Remark} When $N=k$, the cylindric relation can be regarded as a modification of the vanishing 
condition (\ref{vc-Hecke}), although the relation (\ref{vc-Hecke}) is no longer satisfied. However, 
when $n\ge 2$, the vanishing relations (\ref{vc-Hecke}) become cyclic and the cylindric relation 
(\ref{cyclic-rel-N}) is a highly non-trivial relation. 

\subsection{$\check{R}$-matrix}
We give the spin representation of $\check{R}$ introduced in Section~\ref{sec-Hecke}.
The $R$-matrix of the $\mathfrak{gl}_k$ spin chain model is given by
\begin{eqnarray}
R=q\sum_{a=1}^{k}E_{aa}\otimes E_{aa}+\sum_{1\le a\neq b\le k}E_{aa}\otimes E_{bb}
+(q-q^{-1})\sum_{1\le b<a\le k}E_{ab}\otimes E_{ba}. 
\end{eqnarray}
We also introduce the permutation operator $\PP$ and the $q$-permutation operator as follows:
\begin{eqnarray}
\PP&\equiv&\PP_{12}=\PP_{21}=\sum_{a,b=1}^{k}E_{ab}\otimes E_{ba}, \\
\PP^{q}_{12}&=&\sum_{a=1}^{k}E_{aa}\otimes E_{aa}+q\sum_{a<b}^{k}E_{ab}\otimes E_{ba}+q^{-1}\sum_{a>b}^{k}E_{ab}\otimes E_{ba} \nonumber  \\ 
&=&\sum_{a,b=1}^{k}q^{\mathrm{sign}[b-a]}E_{ab}\otimes E_{ba}. \\
(\mathbb{I}\otimes\mathbb{I})_{q}&=&\sum_{a,b=1}^{k}q^{2(a-b)}E_{aa}\otimes E_{bb}.
\end{eqnarray}
The Baxterized $R$-matrix is given by 
\begin{eqnarray}
R_{12}(u)=\frac{a(u)}{\zeta(u)}\PP_{12}+\frac{b(u)}{\zeta(u)}(\mathbb{I}\otimes\mathbb{I}-\PP_{12}^{q})
\end{eqnarray}
where $a(u)=qu^{1/2}-q^{-1}u^{-1/2}$ and $b(u)=u^{1/2}-u^{-1/2}$. Note that $\check{R}=R\PP$.
We also have
\begin{eqnarray}
R_{N1}(u)=a(u)\PP_{N1}+b(z)((\mathbb{I}\otimes\mathbb{I})_{q}-\PP^{q}_{N1})
\end{eqnarray}
corresponding to the affine generator $e_{N}$. The $R$-matrix (and $\check{R}$-matrix) satisfies 
$R_{i+1i+2}=\sigma^{-1}R_{ii+1}\sigma$ for $1\le i\le N-1$ and $R_{NN+1}=R_{N1}$. 

\section{$A_{k}$ Generalized Model on a Cylinder}\label{sec-Akmodel}
We first briefly review the $O(1)$ loop model on a cylinder, which is the $k=2$ case of the 
$A_{k}$ generalized model on a cylinder in Section~\ref{subsec-O1}. We introduce a new model which we call  
$A_{k}$ generalized model on a cylinder in Section~\ref{subsec-Akmodel}. The way of constructing states is given by the 
use of the rhombus tiling.
The relation of our model to the spin chain model is stated in Section~\ref{subsec-Ak-spin}.  
We obtain the sum rule of the $A_{k}$ generalized model by solving the $q$-KZ equation at 
the Razumov-Stroganov point in Section~\ref{subsec-qKZAk}. The solution is identified with a special solution of the 
$q$-KZ equation in Section~\ref{subsec-specialqKZ}.

\subsection{$O(1)$ loop model on a cylinder}\label{subsec-O1}
\subsubsection{The $O(1)$ loop models}
In this subsection, the results of the $O(1)$ loop models are presented. Here is the summary of the 
result if we take $k=2$ in the Section~\ref{subsec-Akmodel}. See also~\cite{DiFZJ04a,DiFZJZu06} for some results and 
details.  

\paragraph{Definition of the $O(1)$ loop models}
The inhomogeneous $O(1)$ loop model is defined on a semi-infinite cylinder of square lattice
with even perimeter $2n$, where squares on the same height are labelled in order cyclically from $1$ to 
$2n$. Spectral parameter $z_{i}$ for $1\le i\le 2n$ is attached to each vertical strip. We attach two kinds 
of unit plaquettes 
\begin{eqnarray}
\scalebox{0.4}{\includegraphics{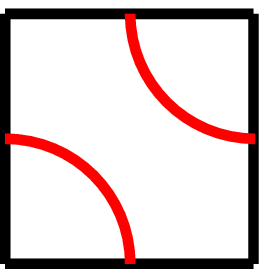}},\ \ \ \ \ \ \scalebox{0.4}{\includegraphics{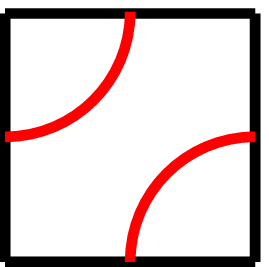}}
\end{eqnarray}
on the square. The weight of the plaquettes in the $i$-th vertical strip is given by the $R$-matrix as
\begin{eqnarray}
R(z_{i},t)=\frac{qz_{i}-q^{-1}t}{qt-q^{-1}z_{i}}\ \raisebox{-0.4\height}{\scalebox{0.36}{\includegraphics{tilea.eps}}}
+\frac{z_{i}-t}{qt-q^{-1}z_{i}}\ \raisebox{-0.4\height}{\scalebox{0.36}{\includegraphics{tileb.eps}}}
\end{eqnarray}
where $t$ is a horizontal spectral parameter. 

\paragraph{States and the boundary condition}
Since red (or grey) lines on a plaquette are non-intersecting, a site is connected to 
another site by a non-intersecting red (or grey) line. From this, all the $2n$ sites are connected to 
each other forming a link. The space of states for the $O(1)$ loop model is 
the set of link patterns. We denote a state by $\pi$, or $|\pi\rangle$.

We introduce the direction of links and the boundary of the cylinder. Let us consider a conformal map from the 
semi-infinite cylinder to a disk with perimeter $2n$. The infinite point is mapped to the origin of the disk. 
The two boundary conditions are classified by whether we regard the origin of the disk as an punctured point 
or not as follows.   
\begin{itemize}
  \item Periodic boundary condition (or unpunctured case) : The infinite point is regarded as an unpunctured 
 point. In this case, we focus only on connectivities between the sites. 
  \item Cylindric boundary condition (or punctured case): The infinite point is regarded as a punctured point. 
 Introducing the punctured point corresponds to introducing a seam between the first and the $2n$-th sites on the 
 cylinder. The direction of a link between sites $i$ and $j$ is measured by $(-1)^{w}$ where $w$ counts how 
 many times the link crosses the seam.   
\end{itemize}

We assign to a loop (even a loop surrounding the punctured point) the weight $\tau=-(q+q^{-1})$ when $q$ is 
a generic value. Note that when $q$ is a cubic root of unity, \textit{i.e.}, $q=-\exp(\pi i/3)$, the weight 
of a loop is $\tau=1$. 

\paragraph{Transfer matrix and $q$-KZ equation}
The row-to-row transfer matrix of the $O(1)$ loop model (in both periodic and cylindric cases) is given by
\begin{eqnarray}
T(t|z_{1},\cdots,z_{2n})=\mathrm{Tr}_{0}\left(
R_{1}(z_{1},t)R_{2}(z_{2},t)\ldots R_{2n}(z_{2n},t)
\right)
\end{eqnarray}
where the trace is taken on the auxiliary quantum space. The transfer matrix naturally acts on a states. 
We want to compute the weight distribution 
$\Psi(z_{1},\cdots,z_{2n})=\sum_{\pi}\Psi_{\pi}|\pi\rangle$ such that 
\begin{eqnarray}
\begin{aligned}
T(t|z_{1},\cdots,z_{2n})\Psi(z_{1},\cdots,z_{2n})=\Psi(z_{1},\cdots,z_{2n}).
\end{aligned}\label{Eigenvec-Psi}
\end{eqnarray}
Instead of the eigenvector problem (\ref{Eigenvec-Psi}), it is enough to consider more generally the 
$q$-KZ equation with two parameters $q$ and $s$ (see~\cite{Pas05,DiFZJ04a})
\begin{eqnarray}\label{qKZ-Psi}
\begin{split}
\check{R}(z_{i},z_{i+1})\Psi(z_{1},\ldots,z_{2n})&=\tau_{i,i+1}\Psi(z_{1},\ldots,z_{2n}), 
 \qquad 1\le i\le 2n-1 \\
\check{R}(z_{2n},sz_{1})\Psi(z_{1},\ldots,z_{2n})&=\Psi(s^{-1}z_{2n},\ldots,sz_{1})
\end{split}
\end{eqnarray}
where $\tau_{i,i+1}$ is an operator acting on a polynomial $f(z_{i},z_{i+1})$ as 
$\tau_{i,i+1}f(z_{i},z_{i+1})=f(z_{i+1},z_{i})$. 
When $s=1$, the eigenvector of the transfer matrix with eigenvalue unity is regarded as 
the solution of the $q$-KZ equation. This is realized at the Razumov-Stroganov point, $i.e.$ $q=-\exp(\pi i/3)$ 
in the link pattern basis~\cite{DiFZJ04a,DiFZJZu06}. The solution of the $q$-KZ equation with generic 
$q$ and $s=q^{6}$ (resp. $s=q^{3}$) was obtained on the link patterns with periodic (rep. cylindric) boundary 
conditions in~\cite{Pas05} (resp. \cite{KasPas06}).

Below, we construct the space of link patterns in the cylindric case, on which the affine 
Temperley-Lieb algebra acts. 

\subsubsection{Word representation (cylindric case)}\label{subsec-TLwords}
In this subsection, the parameter $q$ takes a generic value.
\paragraph{Word representation and cylindric relation}
It is well-known that the word representation of link patterns (periodic case) 
is constructed in the left ideal of the Temperley-Lieb algebra. The lowest state $\omega$ 
is given by the product of $q$-symmetrizer $Y_{1}$, $\omega:=\prod_{i=1}^{n}e_{2i-1}$. 
All the other states are obtained by taking actions of a sequence of $e_{i}$'s, {\it i.e.} words.  

We can construct all the states for the cylindric case in the similar way from $\omega$. 
However, the additional operator $e_{2n}$ appears in the word representation. It is natural that the graphical 
representation of the generators $e_{i}, 1\le i\le 2n-1$, and $e_{2n}$ of the affine Temperley-Lieb algebra is
\begin{eqnarray}
e_{i}=\ \ \raisebox{-0.35\height}{\mbox{\scalebox{0.4}{\includegraphics{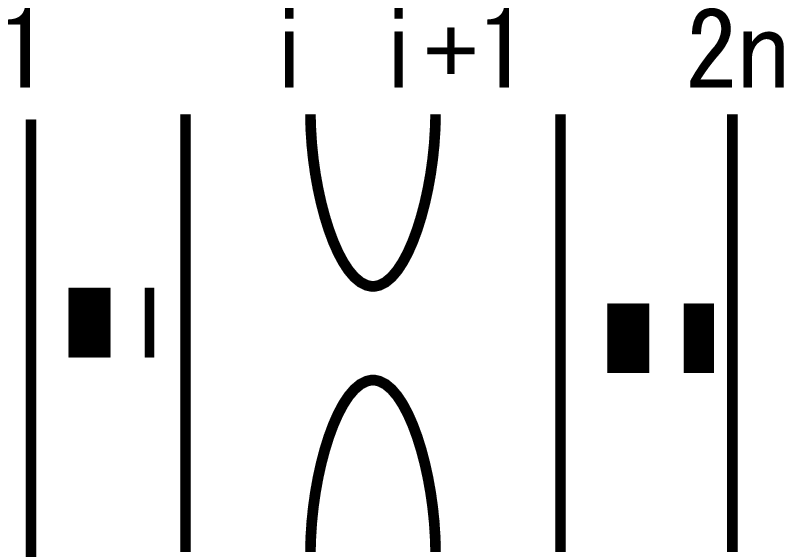}}}}, \qquad
e_{2n}=\ \ \raisebox{-0.35\height}{\mbox{\scalebox{0.4}{\includegraphics{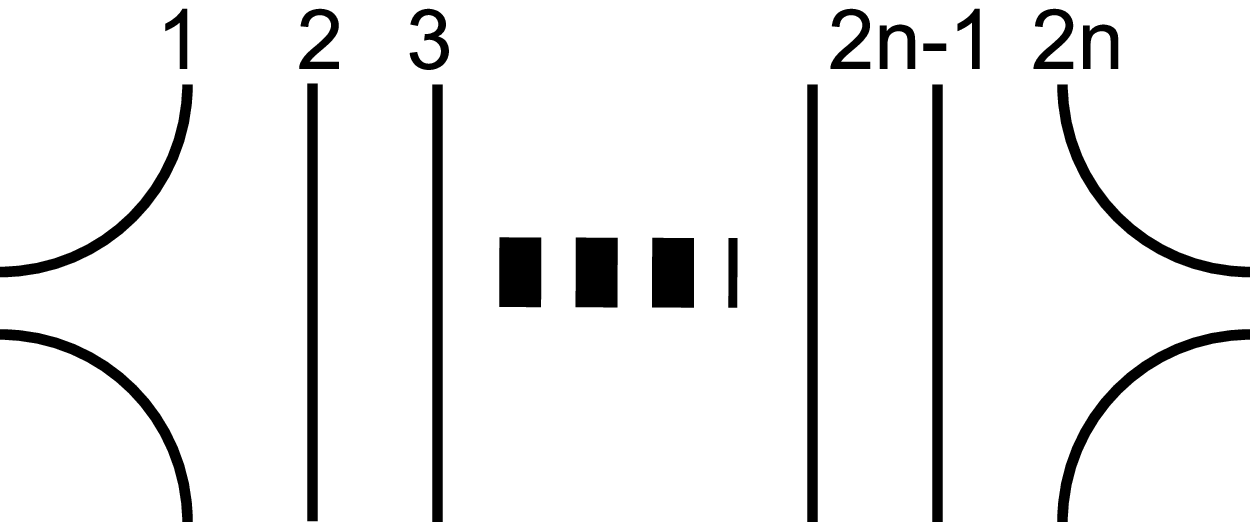}}}}. 
\end{eqnarray}
The cylindric relation (\ref{cyclic-rel-N}) can be written in terms of the affine Temperley-Lieb generators as 
\begin{eqnarray}\label{TL-cyclic-rel}
\prod_{j=1}^{n}e_{2j-1}\prod_{i=1}^{n}e_{2i}\prod_{j=1}^{n}e_{2j-1}=\tau^{2}\prod_{j=1}^{n}e_{2j-1}. 
\end{eqnarray}
This relation can be depicted using the graphical representation as
\begin{eqnarray}\label{TL-cyclic-graph}
\raisebox{-0.55\height}{\mbox{\scalebox{0.4}{\includegraphics{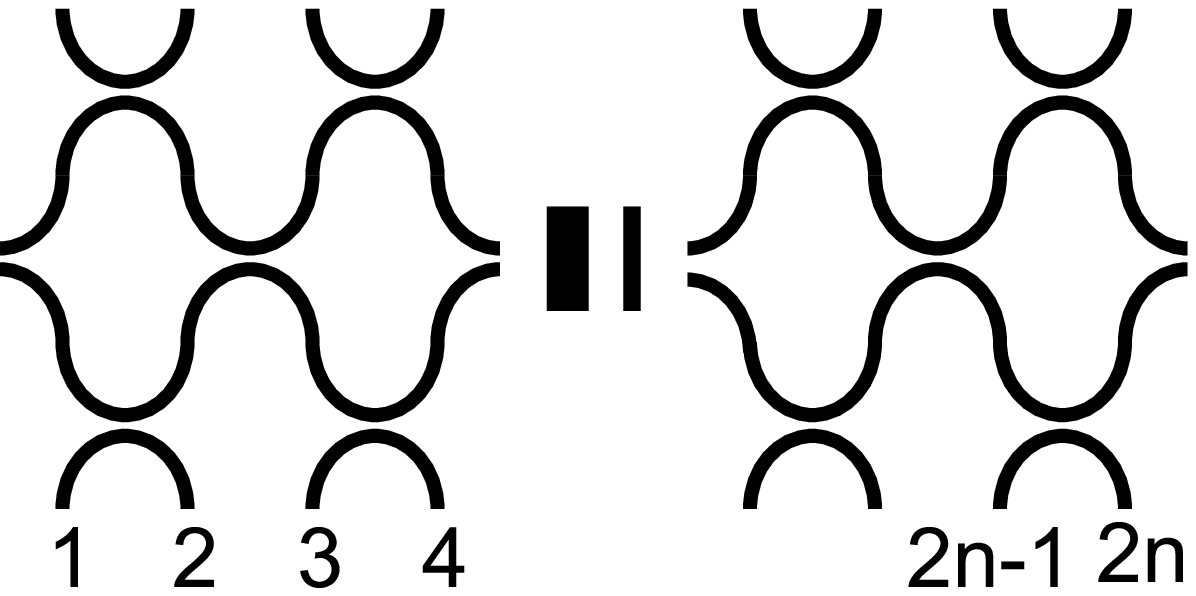}}}} \ \ =\ \tau^{2}\ \ 
\raisebox{-0.5\height}{\mbox{\scalebox{0.4}{\includegraphics{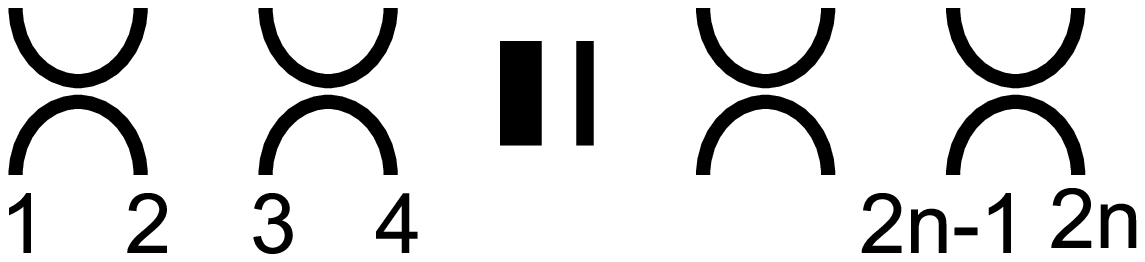}}}}.
\end{eqnarray}
Note that the weight of a loop surrounding the cylinder is $\tau$ and the factor $\tau^{2}$ in 
(\ref{TL-cyclic-rel}) comes from the two loops in the r.h.s. of Eqn.(\ref{TL-cyclic-graph}). 

\paragraph{The highest state}
We write as $\displaystyle \prod_{1\le j\le r}^{\longleftarrow}s_{j}:=s_{r}s_{r-1}\ldots s_{1}$ where 
the order of products is clear, and 
$(s_{r}\ldots s_{1})^{\vee(m)}:=s_{r}\ldots s_{1+(r+m)/2}s_{(r-m)/2}\ldots s_{1}$ for $r>m\ge0$ and 
$r\equiv m\ (\mbox{mod }2)$. We define $(s_{r}\ldots s_{1})^{\vee(m)}=1$ for $m\ge r$ and 
$(s_{r}\ldots s_{1})^{\vee(m)}=s_{r}\ldots s_{1}$ for $m<0$. 
For example, $\displaystyle (\prod_{1\le j\le 5}^{\longleftarrow}s_{j})^{\vee3}=s_{5}s_{1}$ and 
$\displaystyle (\prod_{1\le j\le 6}^{\longleftarrow}s_{j})^{\vee2}=s_{6}s_{5}s_{2}s_{1}$.

The highest state $))\ldots((($ is given in terms of the generators as 
\begin{eqnarray}
\prod_{0\le j\le\left\lfloor n/2\right\rfloor}^{\longleftarrow}\tilde{E}_{j}
\end{eqnarray}
where 
\begin{eqnarray}
\tilde{E}_{j}=\left\{
\begin{array}{cc}
\displaystyle (\prod_{1\le l\le n-1}^{\longleftarrow}e_{2l})^{\vee(2j)}\cdot e_{2n}\cdot  
(\prod_{l=1}^{n}e_{2l-1})^{\vee(2j-1)},  & \mbox{ for $n$ odd} \\
\displaystyle (\prod_{1\le l\le n-1}^{\longleftarrow}e_{2l})^{\vee(2j+1)}\cdot e_{2n}\cdot  
(\prod_{l=1}^{n}e_{2l-1})^{\vee(2j)},  &  \mbox{ for $n$ even} \\
\end{array}\right.
\end{eqnarray}
for $j\ge0$. Other states are obtained as words by acting a sequence of the generators on the highest state. 

\paragraph{\underline{Example: $n=2$}} We have six bases. The correspondence of representation by 
parentheses and by words 
is given as follows: $()()\leftrightarrow e_{1}e_{3}$, $(())\leftrightarrow e_{2}e_{1}e_{3}$, 
$)()(\leftrightarrow e_{2}e_{4}e_{1}e_{3}$, $())(\leftrightarrow e_{1}e_{2}e_{4}e_{1}e_{3}$, 
$)(()\leftrightarrow e_{3}e_{2}e_{4}e_{1}e_{3}$ and $))((\leftrightarrow e_{4}e_{1}e_{3}$. 
Then, the Temperley-Lieb generators are given by \newline
\begin{center}
\scalebox{0.8}{\parbox{15cm}{
\begin{eqnarray}
e_{1}=\left(
  \begin{array}{cccccc}
    \tau  &  1   &  0  & 0   & \tau^{2}   & 1   \\
      0 & 0   & 0   & 0   & 0   & 0   \\
      0 & 0   & 0   & 0  &  0  &  0  \\
      0 & 0   &  1  & \tau & 0   & 0   \\
      0 & 0   & 0   & 0   & 0   &  0  \\
      0 & 0   &  0  &  0  & 0   &   0 \\
  \end{array}
\right), 
e_{2}=\left(
  \begin{array}{cccccc}
    0  &  0 & 0  & 0  & 0   & 0   \\
    1   & \tau   & 0  & 0  & 0  & 0  \\
    0   & 0  & \tau   & 1   & 1   & 1   \\
    0   & 0   & 0   & 0   & 0   & 0   \\
    0   & 0   & 0   & 0   & 0   & 0   \\
    0   & 0   & 0   & 0   & 0   & 0   \\
  \end{array}
\right), 
\sigma=\left(
  \begin{array}{cccccc}
    0  &  0  & \tau   & 0   &  0  & 0   \\
     0  & 0   & 0   &  \tau  & 0   & 0   \\
    \tau^{-1}   &  0  & 0   &  0  &  0  & 0   \\
    0   &  0  &  0  &  0  &  0  & \tau^{-1}   \\
    0   &  \tau^{-1}  & 0   & 0   &  0  &  0  \\
    0   &  0  & 0   &  0  &  \tau  & 0   \\
  \end{array}
\right), 
\end{eqnarray}}} 
\end{center}
and $e_{3}=\sigma e_{2}\sigma^{-1}$ and $e_{4}=\sigma e_{3}\sigma^{-1}$. 
Note that they satisfy the defining Hecke relations and the cylindric relation 
$e_{1}e_{3}e_{2}e_{4}e_{1}e_{3}=\tau^{2}e_{1}e_{3}$.

\paragraph{Relation to the spin chain}
An affine Temperley-Lieb generator $e_{i}\in\mathrm{End}(\mathbb{C}^{2}\otimes\mathbb{C}^{2})$ in 
the spin-$1/2$ representation. This allows us to rewrite a link pattern in terms of the 
spin basis~\cite{MitNiedeGieBat04}. For a given directed link between the site $i$ and $j$ ($i<j$), a spin vector is written as
\begin{eqnarray}
    |\uparrow\downarrow\rangle_{ij}+(-q)^{-1}|\downarrow\uparrow\rangle_{ij},   
    && \ \ \mathrm{for}\  i\rightarrow j, \label{und-link} \\
    |\uparrow\downarrow\rangle_{ij}+(-q)^{-3}|\downarrow\uparrow\rangle_{ij},   
    &&  \ \ \mathrm{for}\ j\rightarrow i. \label{direc-link}
\end{eqnarray}
In the periodic case, every link is expressed as Eqn.(\ref{und-link}) since we do not see the direction. 
In the cylindric case, however, we take a vector of the type (\ref{und-link}) for a link uncrossing 
the seam of the cylinder and a vector of the type (\ref{direc-link}) for a link crossing the seam. 
In both cases, take the tensor product of associated vectors for a link pattern.

\subsubsection{$q$-KZ equation and the sum rule}\label{Sec-qKZ}
The $q$-KZ equation connects the polynomial representation of the affine Temperley-Lieb algebra and the word 
representation of the algebra~\cite{KasPas06,DiFZJZu06}. When $\tau=1$, the $q$-KZ equation can be explicitly solved. 
Then, it is found that the sum rule for $\Psi$ is the product of two Schur functions. When we take the 
homogeneous limit all $z_{i}\rightarrow1$, the sum is proportional to the total number of the $2n\times2n$ 
half-turn symmetric alternating sign matrices (HTSASMs) (see also~\cite{RazStr01a,RazStr01c,BatdeGNie01,Kup02}).

\subsection{$A_{k}$ generalized model}\label{subsec-Akmodel}
We define the $A_{k}$ generalized model on a cylinder. This is the $A_{k}$ generalization of the $O(1)$ loop 
model on a cylinder in Section~\ref{subsec-O1}. This generalization is done by replacing the affine 
Temperley-Lieb algebra and state space labelled by link patterns with the affine Hecke algebra and state 
space labelled by unrestricted paths, respectively. 
In Section~\ref{states-Ak}, we explicitly construct a state by the use of the graphical depiction 
of the $q$-symmetrizer introduced in Section~\ref{subsec-graph-Y}. The relation of our model to the spin chain model is 
discussed in Section~\ref{subsec-Ak-spin}.

We set up the $q$-KZ equation (\ref{qKZ-Psi}) where $\check{R}_{i}(z,w)$ is now the standard 
trigonometric $\check{R}$-matrix defined in (\ref{def-tri-R}). We examine the $q$-KZ equation with two parameters 
$s=1$ and $q=-\exp(\pi i/k+1)$. The sum rule for the solution $\Psi({\bf z})$ is investigated in 
Section~\ref{subsec-qKZAk}.

\subsubsection{States for $A_{k}$ generalized model}\label{states-Ak}
Before constructing the states for the $A_{k}$ generalized model, we introduce some definitions 
and notations. Hereafter, we set $N=nk$. The parameter $q$ is generic in this subsection.  
\begin{definition}\label{def-path}
We define a set of unrestricted paths and restricted paths.
\begin{enumerate}
\item An unrestricted path $\pi:=\pi_{1}\pi_{2}\cdots\pi_{nk}$ of length $nk$ is a set of $nk$ integers satisfying
\begin{eqnarray}
\begin{aligned}
&1\le\pi_{i}\le k \ \ \text{for}\ 1\le i\le nk\\
&\sharp\{i|\pi_{i}=j, 1\le i\le nk\}=n \ \ \text{for}\ \ 1\le j\le k.
\end{aligned}
\end{eqnarray}
\item  If an unrestricted path $\pi$ satisfies 
$\sharp\{i\vert \pi_{i}=j, 1\le i\le l \}\ge\sharp\{i\vert \pi_{i}=j+1, 1\le i\le l \}$ for all $1\le j\le k$ 
and $1\le l\le nk$, the path $\pi$ is said to be a restricted path.
\end{enumerate}
\end{definition}
The number of unrestricted paths is $(nk)!/(n!)^{k}$. The number of restricted paths is the same as the 
number of the standard Young tableaux with shape $k\times n$, {\it i.e.}, 
$(nk)! \prod_{0\le j\le k-1} j!/(n+j)!$.  

A path is graphically depicted by a line graph on a cylinder.  
A line graph $\pi'$ of length $N$ consists of $N+1$ vertex and $N$ edges. The $i$-th and $(i+1)$-th vertices are connected by the $i$-th edge for all $1\le i\le N$. The first and $(N+1)$-th vertices are identified in the case 
of a line graph on a cylinder. When the angular coefficient of the $i$-th edge is $\pi(k-2m+1)/2k$ 
with $1\le m\le k$, the $i$-th edge is said to be of type $m$. A path $\pi$ is identified with a line graph  of
length $nk$ on a cylinder where the $i$-th edge is of type $\pi_{i}$.

\bigskip
In Section~\ref{subsec-graph-Y}, we have seen that a $q$-symmetrizer $Y_{k}$ corresponds to 
a $2(k+1)$-gon. Recall that $Y_{q\mbox{-sym}}$ is the product of the $q$-symmetrizers.
\begin{definition}\label{def-graph-omega}
The graphical representation of $Y_{q\mbox{\rm -sym}}=\prod_{i=0}^{n-1}Y_{k}(e_{ik+1},\ldots,e_{(i+1)k})$  is  
the graph where we put $n$ $2(k+1)$-gons side by side. The terminal vertices are identified as the graph is on a cylinder. 
\end{definition}

\noindent
{\bf Example} The graphical expression of $Y_{q\mbox{-sym}}$ is shown for $k=6$. The cylinder is cut along the dotted 
line.  
\begin{eqnarray}
Y_{q\mbox{-sym}}=\raisebox{-0.3\height}{\scalebox{0.8}{\includegraphics{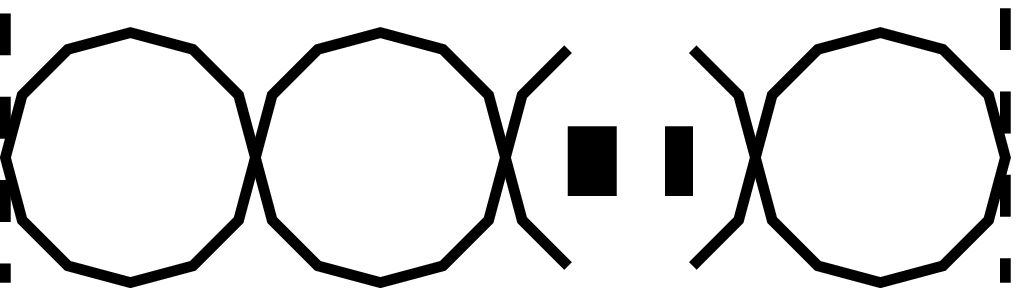}}}. 
\end{eqnarray}
\bigskip

We have $nk$ edges on the top of the graphical representation of $Y_{q\mbox{-sym}}$. 
From Definition~\ref{def-path} and Definition~\ref{def-graph-omega}, we have a corresponding 
path as follows.
\begin{proposition}
The top edges of $Y_{q\mbox{\rm -sym}}$ are identified with the path $\pi^{\Omega}$ where $\pi^{\Omega}_{pk+q}=q$
for $0\le p\le n-1$ and $1\le q\le k$. 
\end{proposition}

Consider a path $\pi$ that satisfies $\pi_{i}>\pi_{i+1}$ for a given $i$. In this situation, we can pile a rhombus with a 
positive integer $m$, corresponding to $\check{L}_{i}(m)$, locally over the line graph consisting of the $i$-th and 
the $(i+1)$-th edges. Then, we obtain a new path $\pi'$ satisfying 
\begin{eqnarray}
\begin{aligned}
&\pi^{'}_{j}=\pi_{j}\ \ \ \text{for}\ j\neq i,i+1\\
&\pi^{'}_{i}=\pi_{i+1},\ \ \pi^{'}_{i+1}=\pi_{i}.
\end{aligned}\label{path-tiling}
\end{eqnarray}
where $\pi_{kn+1}=\pi_{1}$. We introduce an order of paths such that if (\ref{path-tiling}) is satisfied  
the path $\pi^{'}$ is lower than $\pi$. The order of paths is the same as the one of tiling rhombi. 
 
\begin{definition}
We pile rhombi from bottom to top over the graphical representation of $Y_{q\mbox{-sym}}$, or equivalently,
over the path $\pi^{\Omega}$. We pile rhombi one by one following the rule in Eqn.(\ref{path-tiling}).
\end{definition}
From the above definition, we have a natural map from a rhombus tiling over $\pi^{\Omega}$ to an unrestricted 
path. The top edges of the rhombus tiling are identified with a line graph on a cylinder and a path. 

\begin{definition} \label{def-word}
We introduce a word and a reduced word.
\begin{enumerate}
\item Let $l$ be a positive integer. $M=\{m_{j}\vert 1\le m_{j}\le k-1, 1\le j\le N\}$ and 
$I=\{i_{j}\vert 1\le i_{j}\le N, 1\le j\le l\}$ are sets of positive integers.
A word $w$ of length $l$ is defined as 
\begin{eqnarray}\label{def-L-word}
w=\check{L}_{i_{1}}(m_{1})\ldots\check{L}_{i_{l}}(m_{l})Y_{q\mbox{\rm -sym}}
\end{eqnarray}
for some $\{l,M,I\}$ and $Y_{q\mbox{-sym}}=\prod_{i=0}^{n-1}Y_{k}^{(ik+1)}$. By definition, 
$Y_{q\mbox{-sym}}$ is itself a word of length zero. The length of a linear combination of words is identified by the 
maximum length of words in it.
\item A word $w$ is said to be equivalent to another word $w'$ if we obtain $w$ from $w'$ only 
by using the defining relations of the affine Hecke algebra, (\ref{Hecke-relation}), (\ref{vc-Hecke}) and the 
cylindric relations (\ref{cyclic-1}) or (\ref{cyclic-general}).
\item A word $w$ (of length $l$) is said to be a reduced word if there exists no equivalent 
word of length $l'$ with $l'<l$. 
\end{enumerate}
\end{definition}
Hereafter, a word means a reduced word. 

The word representation is the representation of the affine Hecke algebra on the left ideal 
$\widehat{H_{N}^{(k)}}Y_{q\mbox{-sym}}$. 

\noindent
{\bf Remark} We restrict ourselves to $1\le m_{j}\le k-1$ in Definition \ref{def-word}. Since 
$\check{L}_{i}(m)=\check{L}_{i}(m-1)+(\mu_{m-1}-\mu_{m})$, a word can be rewritten in terms 
of other words. The vanishing condition (\ref{vc-Hecke}) and the graphical representation of a $q$-symmetrizer 
in Section~\ref{subsec-graph-Y} imply that $1\le m_{j}\le k-1$ is enough to have non-vanishing words. 

From these definitions, we have the following map from a rhombus tiling to a word.
For a given rhombus tiling over $\pi^{\Omega}$, we have a natural map
from a rhombus tiling with integers to a word $w$ as Eqn.(\ref{def-L-word}).
\bigskip

The above definitions and proposition are summarized as follows. We have a word and an unrestricted path 
$\pi$ for a rhombus tiling with integers. There are, however, many rhombus tilings with integers whose top 
edges are characterized by the path $\pi$, whereas we have only one word for a given rhombus tiling 
with integers.

\bigskip

We want to get a state $|\pi\rangle$ labelled by an unrestricted path $\pi$ satisfying 
the following properties.
\begin{enumerate}[(P1)]
  \item\label{P1}If $\pi$ satisfies $\pi_{i}>\pi_{i+1}$, the state is invariant under 
  under the action of $e_{i}$, \textit{i.e.}  $e_{i}|\pi\rangle=\tau|\pi\rangle$.   
  \item\label{P2} If $\pi$ satisfies $\pi_{i}<\pi_{i+1}$, the action of $e_{i}$ is given by  
  $e_{i}|\pi\rangle=\sum_{\pi'}C_{i,\pi,\pi'}|\pi'\rangle$. If the coefficient 
  $C_{i,\pi,\pi'}\neq0$, $\pi'$ is obtained by adding a unit rhombus as (\ref{path-tiling})  or a path below $\pi$. 
  \item\label{P3} If $\pi$ satisfies $\pi_{i}=\pi_{i+1}$, the action of $e_{i}$ is given 
  by $e_{i}|\pi\rangle=\sum_{\pi'}C_{i,\pi,\pi'}|\pi'\rangle$. If the coefficient 
  $C_{i,\pi,\pi'}\neq0$, $\pi'$ is a path below $\pi$.
  \item\label{P4} In the properties (P\ref{P2}) and (P\ref{P3}), let us consider the case where 
  $e_{j}|\pi\rangle=\tau|\pi\rangle$ for $j\neq i\pm1$. Then, for a path $\pi'$ with non-zero $C_{i,\pi,\pi'}$ 
  it satisfies $e_{j}|\pi'\rangle=\tau|\pi'\rangle$. 
\end{enumerate}

We will have the one-to-one correspondence;
\begin{eqnarray}
{\mbox{a state $|\pi\rangle$}\atop {\mbox{labelled by a path $\pi$}}} \Longleftrightarrow
{\mbox{a rhombus tiling} \atop {\mbox{with integers}}} \Longleftrightarrow
{\mbox{a word}}.
\end{eqnarray}
A specific choice of a rhombus tiling for a given path $\pi$ allows us to describe the state $|\pi\rangle$ 
in terms of a word. 
The difficulty is to assign positive integers to rhombi for a given rhombus tiling. 

First of all, 
\begin{definition}\label{def-state-omega}
The word representation of the state $|\pi^{\Omega}\rangle$ is $Y_{q\mbox{-sym}}$. 
The graphical representaion of $|\pi^{\Omega}\rangle$ is the graph where we put $n$ $(2k+1)$-gon 
with integers side by side. 
\end{definition}

Before constructing enery state $|\pi\rangle$, we prepare some terminologies and notations.
Let us start to assign integers to every corners of rhombi as follows (see also the explanation below 
Eqn.(\ref{L-rhombus})).  
\begin{definition}
Consider a rhombus with an integer $m$. We assign  $+m$ (resp. $-m$) to up and down (resp. right and left) 
corners of the rhombus. 
\end{definition}

\begin{definition}
Suppose that a vertex is shared by some rhombi. We assign to the vertex the sum of all the integers on 
corners around the vertex. 
\end{definition}
\begin{definition}[zero-sum rule]
Suppose that a vertex is completely surrounded by rhombi. A vertex is said to satisfy the zero-sum rule if
the sum of all signed integers on corners surrounding the vertex is equal to zero.  
\end{definition}
Note that all the vertices inside $2(k+1)$-gon of a $q$-symmetrizer $Y_{k}$ satisfy the zero-sum 
rule and that the vertices on the top path have integer one.

\begin{definition}
Fix an integer $1\le l\le k$. 
Consider $l$ integers $1\le i_{1}<i_{2}<\cdots<i_{l}\le k$. Let 
$b^{top}$ be a partial path of length $l$ satisfying $b^{top}_{j}=i_{j}$ for all $1\le j\le l$. Fix 
a path of length $l$, $b^{bot}$, which satisfies each $i_{j}$ appears once in 
$\{b^{bot}_{i}\}$, $b^{bot}_{1}\neq i_{1}$ and $b^{bot}_{l}\neq i_{l}$. Connect $i$-th and 
$(i+l)$-th vertices by the two line graphs of $b^{top}$ and $b^{bot}$. The $2l$-gon surrounded by the two 
line graphs is called a rhombus block $B_{i,i+l}$ surrounded by $b^{top}$ and $b^{bot}$. 
\end{definition}

\begin{proposition}\label{numbering-block}
A positive integer on every rhombus in a given rhombus block $B_{i,i+l}$ is uniquely determined by 
the zero-sum rule and the integers on the vertices, from the $(i+1)$-th to the $(i+l-1)$-th vertices in the 
top partial path $b^{top}$. 
\end{proposition}
\begin{proof}
Adding some rhombi to $B_{i,i+l}$ to form a $2l$-gon looking like a $q$-symmetrizer, it is sufficient to show 
that positive integers on rhombi in the $2l$-gon are uniquely determined by the condition for $B_{i,i+l}$. 
If we change the rhombus tiling of the $q$-symmetrizer by elementary moves, we have the form of 
the standard rhombus tiling of it as shown in Fig.~\ref{graphical-q-sym-eps} but integers are different. 
Since two edges $b^{top}_{i}$ and $b^{top}_{i+1}$ form a rhombus at the $(i+1)$-th vertex, the integer 
on the rhombus is the same as the integer on the $(i+1)$-th vertex in $b^{top}$. The integer on the next rhombus 
is determined by the integers on the first rhombus and on the $(i+2)$-th vertex in $b^{top}$. Integers 
on $l-1$ rhombi whose edge is the part of the top partial path $b^{top}$ are determined one-by-one in this way. 
Other remained rhombi form a smaller polygon , {\it i.e.}, $2(l-1)$-gon. All the integers for this $2(l-1)$-gon 
are fixed by the zero-sum rule. Next, by elementary moves we get back to the equivalent expressions of $2l$-gon 
with integers, one of which contains the rhombus block $B_{i,i+l}$.  By taking away certain pieces of rhombi 
from $2l$-gon, we obtain the block $B_{i,i+l}$ with integers. 
\end{proof}

Recall that we may have many ways of rhombus tilings with integers corresponding to a path $\pi$. 
All the unrestricted paths are obtained up to a certain finite height starting from the lowest path 
$\pi^{\Omega}$. For a path we take a rhombus tilings with the smallest number of rhombi. However, we 
have many equivalent rhombus tilings because of elementary moves of rhombi. 

In particular, we want a rhombus tiling representing a state $|\pi\rangle$ satisfying the properties from 
(P\ref{P1}) to (P\ref{P4}). This is possible by using the freedom by elementary moves of rhombi. 

Now, we explain the construction of a state $|\pi\rangle$ for a given path $\pi$. The procedure is divided 
into three steps. First, we fix a rhombus tiling for a given path. We divide the rhombus tiling into 
some pieces of rhombus blocks. Secondly, we assign integers on all the rhombi for the rhombus tiling.  
Finally, we identify the state $|\pi\rangle$ with one of rhombus tiling with integers. 

\bigskip
\noindent
{\bf Step1:}
Take one of its rhombus tilings which gives a path $\pi$. Fix an order of tiling rhombi from bottom. 
We have a set of lower paths than $\pi$ 
associated with this tiling. Here, we divide a given rhombus tiling into pieces of rhombus blocks. 
\begin{enumerate}[{Step1}-1]
  \item Consider a convex partial path $\pi_{i,i+m}:=\pi_{i}\ldots\pi_{i+m}$ in $\pi$ satisfying 
  $\pi_{i-1}>\pi_{i}<\ldots<\pi_{i+m}>\pi_{i+m+1}$ for some $m\ge 1$. Take a lower convex 
  partial path $\pi'$ as long as possible such that $\pi'$ contains the partial path $\pi_{i,i+m}$.  We 
  call $\pi'$ the longest convex sequence (lc-sequence) associated with $\pi_{i,i+m}$.  
  Write down all the longest convex sequences for the path $\pi$.
 
  \item If two lc-sequences cross at a vertex  below the path $\pi$, we modify them as follows. We keep 
  the longer lc-sequence as it is. We split the shorter lc-sequence into two parts at the crossing point, and 
  take away the part beneath the longer one.  See Fig.~\ref{fig:path-shorten}. 
  When two crossing sequences have the same length, one of these are to be shortened in the similar way.
\begin{figure}[htbp]
  \begin{center}
  \scalebox{0.8}{\includegraphics{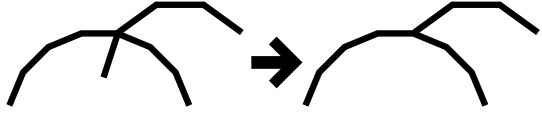}}
  \caption{A part of a lc-sequence beneath a longer lc-sequence is taken away.}
  \end{center}  \label{fig:path-shorten}
\end{figure}

  \item \label{step} Let us denote by $\pi'_{i,i+l}=\pi'_{i}\pi'_{i+1}\cdots\pi'_{i+l-1}$ the lc-sequence from the $i$-th  
  to the $(i+l)$-th vertex, which contains two edges of the last piled rhombus. Let ${\pi'}_{i,i+l}^{low}$ be a partial 
  lower path from from the $i$-th vertex to $(i+l)$-th as low as possible. Then, we have a rhombus block $B_{i,i+l}$
  surrounded by $\pi'_{i,i+l}$ and ${\pi'}_{i,i+l}^{low}$. If there is a part of another lc-sequence $\pi_{lc}$ 
  inside the block $B_{i,i+l}$, the lowest path $\pi_{i,i+l}^{low}$ is to be modified such that $\pi_{i,i+l}^{low}$
  is over $\pi_{lc}$.  
  
  \item Successively, take away rhombus blocks obtained in Step1-\ref{step}.  Finally, we have the 
  path $\pi^\Omega$ and many pieces of rhombus block.
\end{enumerate}
\begin{flushright}
(Step1 ends)
\end{flushright}
We make an order of piling rhombus blocks as follows. 
If two blocks are far enough, we may exchange the order of two blocks. 
However, if $\sharp\{ k\vert i\le k\le i+l$, $j\le k\le j+l^{'} \}\ge1$ for given two blocks $B_{i,i+l}$ 
and $B_{j,j+l^{'}}$, the order of two blocks is determined by the order of piling rhombi. 
Below, we  fix an order of piling rhombus blocks. The order of removing rhombus blocks is the reverse 
of the piling one.

\bigskip
\noindent
{\bf Step2:}
We are ready to assign positive integers to all rhombi for a given rhombus tiling. 
\begin{enumerate}[{Step2-}1]
  \item Let us consider the first rhombus block in the removing order. We assign the positive 
  integer $1$ to all convex vertices on the top partial path of the block. From 
  Proposition~\ref{numbering-block}, we determine integers on all rhombi in this block.  

  \item We move to the second rhombus block. If an integer on a convex vertex in the top partial path of 
  the second block is determined from the integers on the first block by the zero-sum rule, we assign that 
  integer on the vertex. Otherwise, we assign $1$ on them. Again, we determine integers on all rhombi in the 
  second block from Proposition~\ref{numbering-block}. 
 
  \item We determine integers on all subsequent rhombus blocks in the similar way. We continue this process 
  until we assign integers on the all rhombi over $\pi^{\Omega}$. 
\end{enumerate}
\begin{flushright}
(Step2 ends)
\end{flushright}
From the construction, all the vertices inside the rhombus blocks satisfy the zero-sum rule. 
All the integers on convex vertices in the top path of $Y_{q\mbox{-sym}}$ are one. However, the zero-sum rule may 
not hold on vertices in the top path of $Y_{q\mbox{-sym}}$. This is because there is a concave vertex on the 
top path. 
\bigskip

\noindent
{\bf Step3:}
We fix integers for a given path $\pi$ and its rhombus tiling with integers. Then, we choose one of rhombus 
tilings with integers as a state $|\pi\rangle$. 

We introduce a sequence of integers $\mu=(\mu_{1},\ldots,\mu_{k})$ for a rhombus tiling $\mu$ where 
$\mu_{j}$ is the total number of positive integer $j$ written in rhombi. Let $\mu$ and $\nu$ be rhombus 
tilings corresponding to the same path, then we may have the natural order. $\mu\succeq\nu$ means 
$\mu_{k}>\nu_{k}$, or $\mu_{k-r}=\nu_{k-r}$ for all $0\le r\le i-1$ and $\mu_{k-i}>\nu_{k-i}$ for some 
$1\le i\le k$, and $\mu=\nu$ holds when $\mu_{j}=\nu_{j}$ for all $1\le j\le k$. 

\begin{definition}
We choose a rhombus tiling with $\mu$ for a path $\pi$ such that $\mu\succeq\nu$ for any other 
rhombus tiling with $\nu$. The state $|\pi\rangle$ is identified  with the word of rhombus tiling 
with $\mu$.  
\end{definition}
\begin{flushright}
(Step$3$ ends)
\end{flushright}

Some remarks are in order. 

\noindent
{\bf Remark1} The cylindric relation is expressed by means of a rhombus tiling with integers.
Rewrite the cylindric relations as 
\begin{eqnarray}
Y_{k-1}(e_{1},\cdots,e_{k-1})\check{L}_{k}(k-1)Y_{k-1}(e_{1},\cdots,e_{k-1})
=\mu_{k}^{-1}\alpha_{k-1}Y_{k-1}(e_{1},\cdots,e_{k-1})
\end{eqnarray}
for the case where $N=k$. Here, $\alpha_{k}$ is given by Eqn.(\ref{eigenvalue-Y}). And 
\begin{eqnarray}\label{q-sym-for-graph}
Y_{q\mbox{-sym}}\cdot\left(\prod_{i=1}^{n}\check{L}_{ik}(k-1)\right)\cdot Y_{q\mbox{-sym}}
=\Delta_{k}^{n-1}\mu_{k}^{-1}\alpha_{k-1}^{n}Y_{q\mbox{-sym}}
\end{eqnarray}
for the case where $N=nk$ with $n\ge2$. Here, $\Delta_{k}=\mu_{k}-\mu_{k-1}$. The graphical representation of these, 
which is the generalization of Eqn.(\ref{TL-cyclic-graph}), indicates that it is possible to truncate piles of 
rhombi by some height. 

\noindent
{\bf Example:} $k=6, n\ge2$. We have the following graphical representation.
\begin{eqnarray*}
\raisebox{-0.4\height}{\scalebox{0.6}{\includegraphics{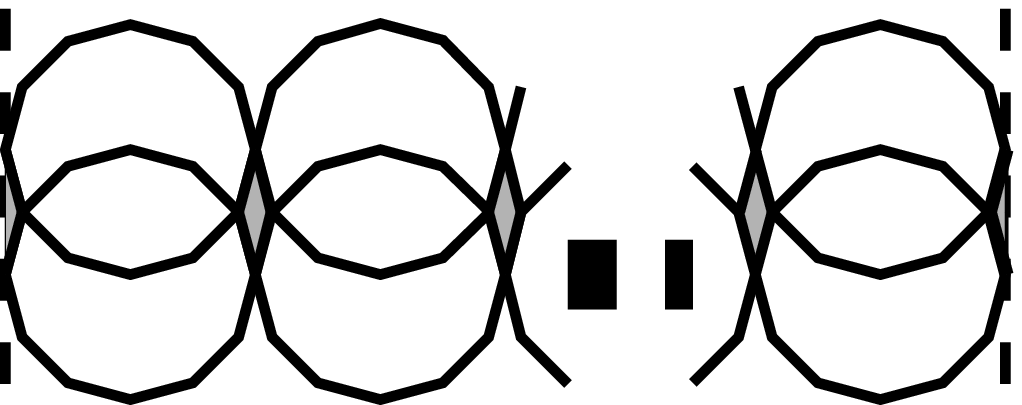}}}\quad=C_{n}\quad
\raisebox{-0.4\height}{\scalebox{0.6}{\includegraphics{left_ideal.eps}}}
\end{eqnarray*}
where $C_{n}=\Delta_{6}^{n-1}\mu_{6}^{-1}\alpha_{5}^{n}$, we put $5$ on the grey rhombi, and the  
dodecagons are the $q$-symmetrizers $Y_{5}$. Polygons are supposed to be covered by rhombi with integers. 
We choose one of the equivalent expressions of the $q$-symmetrizer such that two dodecagons share 
the same rhombus of tiling of intersectional octagon with integers. The rhombus for the operator 
$\check{L}_{N}$ are divided into two parts since the cylinder is cut along the dotted line. 

\bigskip
\noindent
{\bf Remark2:} We briefly explain states constructed in the above way satisfy the properties from (P\ref{P1}) 
to (P\ref{P4}). When an integer on a vertex in the path $\pi$ is one (the vertex is convex by definition), we can bring 
the rhombus with one, $\check{L}_{i}(1)=e_{i}$, to the convex vertex from somewhere inside the rhombus tiling 
by the elementary moves of rhombi. From the first relation in Eqn.(\ref{Hecke-relation}), we have the property (P\ref{P1}).

Notice that $e_{i}=\check{L}_{i}(1)$ and $\check{L}_{i}(m)=\check{L}_{i}(m-1)+(\mu_{m-1}-\mu_{m})$. The action 
of $e_{i}$ on $|\pi\rangle$ leads to getting the state $\pi'$ by (\ref{path-tiling}) and other states 
for lower paths. We have the property (P\ref{P2}).  

Consider a path $\pi$ and suppose that (P\ref{P3}) holds true for all lower states than $\pi$. When $\pi$ 
satisfies $\pi_{i}=\pi_{i+1}$, take an action of $e_{i}$ on the word for the state $|\pi\rangle$. By using the 
relation $e_{i}e_{j}=e_{j}e_{i}$ for $|i-j|\ge2$, we change the position of $e_{i}$ from left to right as 
many times as possible. Then we have a word $\mathcal{L}e_{i}\mathcal{L}' Y_{q\mbox{-sym}}$ where 
$\mathcal{L},\mathcal{L}'$ are sequences of $\check{L}$. $\mathcal{L}$ does not include $e_{j}$, $j=i,i\pm1$ and $\mathcal{L}'=\check{L}_{i\pm1}\cdots$. The word $e_{i}\mathcal{L}'Y_{q\mbox{-sym}}$ is written in terms of 
other words whose paths are lower than that of $\mathcal{L}'Y_{q\mbox{-sym}}$. When $\mathcal{L}$ is the 
identity, we apply the similar argument for $e_{l}$ with a certain $l$ instead of $e_{j}$ by successively 
using the Hecke relation $e_{i}e_{i\pm1}e_{i}-e_{i}=e_{i\pm1}e_{i}e_{i\pm1}-e_{i\pm1}$. From the assumption, 
$e_{i}\mathcal{L}' Y_{q\mbox{-sym}}$ is a linear combination of the lower states. After acting $\mathcal{L}$ on 
the obtained states, we get $e_{i}\mathcal{L}\mathcal{L}'Y_{q\mbox{-sym}}$ as a linear combination of 
the lower states $|\pi\rangle$. 

Finally, the property (P\ref{P4}). Let $|\pi\rangle$ satisfy $e_{j}|\pi\rangle=\tau|\pi\rangle$. In the rhombus 
tiling, we may pile the rhombus for $\check{L}_{j}(1)$ at the last. In all the cases 
$\pi_{i}\gtreqqless\pi_{i+1}$ for $i\neq j,j\pm1$, all the words in the word expansion of 
$e_{i}|\pi\rangle$ has the property that $e_{j}$ can be moved the leftmost. (P\ref{P4}) follows from this observation.

\bigskip
\noindent
{\bf Remark3:} The cyclic operator $\sigma$ acts on a state $|\pi\rangle$ as follows. Let us introduce 
the cyclic operator $\bar{\sigma}$ acting on an unrestricted path as 
$\bar{\sigma}:\pi\mapsto\pi'=\pi_{2}\ldots\pi_{N}\pi_{1}$. The action of $\sigma$ is given by 
\begin{eqnarray}\label{cyclic-word}
\sigma|\pi\rangle=C_{\bar{\sigma}\pi}|\bar{\sigma}\pi\rangle
\end{eqnarray}
with a certain constant $C_{\bar{\sigma}\pi}$. From $\bar{\sigma}^{N}=1$, we normalize $\sigma^{N}=1$, or 
equivalently, $\prod_{k=1}^{N}C_{\bar{\sigma}^{k}\pi}=1$. To see why Eqn.(\ref{cyclic-word}) works, consider 
the action of $\sigma$ on the state $|\pi^\Omega\rangle$. Note that 
$\sigma e_{i}|\pi^\Omega\rangle=e_{i-1}\sigma|\pi^\Omega\rangle$ and 
$e_{pk+q}|\pi^\Omega\rangle=\tau|\pi^\Omega\rangle$ holds for $0\le p\le n-1$ and $1\le q\le k-1$. 
The state $|\bar{\sigma}\pi^\Omega\rangle$ is the only state that has the same convexity as 
$\sigma|\pi^\Omega\rangle$. Then we have $\sigma|\pi^\Omega\rangle\propto |\bar{\sigma}\pi^{\Omega}\rangle$. 
The action of $\sigma$ on a state $|\pi\rangle=\prod_{i\in I}\check{L}_{i}(m_{i})|\pi^{\Omega}\rangle$ is 
written as 
$\sigma|\pi\rangle=\prod_{i}\check{L}_{i-1}(m_{i})\sigma|\pi^\Omega\rangle\propto|\bar{\sigma}\pi\rangle$. 
Eqn.(\ref{cyclic-word}) follows from these considerations.

\bigskip
\noindent
{\bf Remark4:}
Although we are dealing with the affine Hecke algebra considered in Section~\ref{sec-Hecke}, most of the above 
statements are also available to the Hecke algebra just by reducing the state space to only restricted paths. 
The cyclic operator is written in terms of the Hecke generators as $\sigma=t_{N-1}^{-1}\ldots t_{1}^{-1}$. 
In other words, the vector space spanned by states labelled by unrestricted paths is reducible in the sense of 
Definition~\ref{def-word} if we consider the Hecke algebra.

We have no relation like the cylindric relations for the case of the Hecke algebra and do not have the rhombus for 
$\check{L}_{N}$ since the algebra has no affine generator $e_{N}$. 
Afterall, the piling of rhombi stops when we have the path $1_{n}2_{n}\cdots k_{n}$ where 
$l_{n}=\underbrace{l\cdots l}_{n}$. The case of restricted paths is considered in \cite{DiFZJ05a}.

\subsubsection{Relation to spin chain model}\label{subsec-Ak-spin}
From the above construction of states of the $A_{k}$ generalized model, we see a bridge between the $A_{k}$ model 
and the spin chain model. 

The spin representation of the generators of the affine Hecke algebra gives the $R$-matrix of the spin chain 
as in~\cite{Bax82,WadDegAku89,Jim86}. The transfer matrix of the $U_{q}(\mathfrak{gl}_{k})$ spin chain is 
\begin{eqnarray}
T(u)=\mathrm{Tr}_{0}R_{01}(u)R_{02}(u)\cdots R_{0L}(u),
\end{eqnarray}
where the trace is taken for the auxiliary quantum space indexed by $0$. The Hamiltonian is given by~\cite{Jim86}
\begin{eqnarray}
\mathcal{H}=T^{-1}(u)\frac{d T(u)}{du}\Big|_{u=1}=\sum_{i=1}^{L-1}\mathcal{H}_{i,i+1}+\mathcal{H}_{L,1},
\end{eqnarray}
where 
\begin{eqnarray}
\mathcal{H}_{i,i+1}&=&\frac{1}{q-q^{-1}}\mathcal{P}_{i,i+1}R_{i,i+1}^{'}\ \ \ 1\le i\le L-1, \\
\mathcal{H}_{L,1}&=&\frac{1}{q-q^{-1}}\mathcal{P}_{L,1}\tilde{R}_{L,1}^{'}, \\
R^{'}_{ab}&=&2(\mathbb{I}\otimes\mathbb{I}-\mathcal{P}_{ab}^{q})+(q+q^{-1})\mathcal{P}_{ab}, \\
\tilde{R}_{ab}^{'}&=&2((\mathbb{I}\otimes\mathbb{I})_{q}-\mathcal{P}_{ab}^{q})+(q+q^{-1})\mathcal{P}_{ab}.
\end{eqnarray}
where $\mathcal{P},\mathcal{P}^{q}$ and $\tilde{\mathcal{P}}^{q}$ are permutations defined in 
Section~\ref{sec-Spin}. We focus on the eigenvector of the Hamiltonian $\mathcal{H}$. 

Suppose that $\Psi({\bf z})$ is the solution of the $q$-KZ equation~(\ref{qKZ-Psi}) at $q=-\exp(\pi i/(k+1))$. 
Because of the commutation relation between the transfer matrix of 
the $A_{k}$ generalized model and that of the spin chain, the solution $\Psi$ in the homogeneous limit is 
also the eigenvector of the spin chain at $q=-\exp(\pi i/(k+1))$. 
Spin chain models have nice properties at this special point. For instance, it is conjectured that 
the free-fermion part of the spectrum of the $SU_{q}(k+1)$ Perk-Schultz at $q=-\exp(\pi i/(k+1))$  
is a consequence of nice properties of the inhomogeneous $SU_{q}(k)$ vertex model~\cite{AlcStro03}.

Let us discuss the relation between the states of the $A_{k}$ generalized model and the vector space of the spin 
chain model. All the states of the $A_{k}$ model are obtained by acting a sequence of $e_{i}$'s on the product 
of the $q$-symmetrizer, $Y_{q\mbox{-sym}}$. From Proposition~\ref{eigenvec-Y} in the spin representation, 
$|v_{0}\rangle$ (Eqn.\ref{spin-eigenvec-Y}) is the only eigenvector of $Y_{q\mbox{-sym}}$ with non-zero
eigenvalue. Together with Eqn.(\ref{relation-Y-v}), the lowest state $|\pi^{\Omega}\rangle$ is identified 
with the vector $|v_{0}\rangle$ in the spin representation. All the other states are expressed in terms of 
vectors in the spin representation by multiplying  $|v_{0}\rangle$ by a sequence of $\check{L}$-matrix in the 
spin representation. This is a natural generalization of Eqn.(\ref{und-link}) and 
Eqn.(\ref{direc-link}). We are not able to write down the expression as simply as in the case of the $XXZ$ spin 
chain. However, this allows us to write down the solution $\Psi({\bf z})$ in terms of the spin representation. 

The spin chain considered here is the model of spin-$(k-1)/2$. 
The total number of sites $n_{i}$ with $S_{z}=(2i-k-1)/2$, $1\le i\le k$ are conserved quantities. 
The state space of the $A_{k}$ generalized model is the vector space
of spin vectors with all $n_{i}=n$ in the spin representation. We can analyze the $A_{k}$ generalized model in the 
spin representation by constructing words.

\subsubsection{$q$-KZ equation and the sum rule}\label{subsec-qKZAk}
We solve the $q$-KZ equation following the method used in~\cite{DiFZJ04a,DiFZJZu06}. The solution is supposed to be 
the one of the minimal degree. In particular, we consider the solution 
with $q$ a root of unity, $q=-\exp(\pi i/(k+1))$ and $s=1$. 
 
\paragraph{$q$-KZ equation}
The $q$-KZ equation (\ref{qKZ-Psi}) at $(s,q)=(1,-\exp(\pi i/(k+1)))$ is rewritten as 
\begin{eqnarray}\label{qKZ-t-e}
t_{i}\Psi({\bf z})=(e_{i}-\tau)\Psi({\bf z}),\qquad 1\le i\le N
\end{eqnarray}
where $z_{N+1}=z_{1}$ and $t_{i}$ acts on a polynomial $f({\bf z}):=f(z_{1},\ldots,z_{N})$  as
\begin{eqnarray}
t_{i}f({\bf z}):=\frac{qz_{i}-q^{-1}z_{i+1}}{z_{i+1}-z_{i}}(\tau_{i}-1)f({\bf z})
\end{eqnarray}
and $\tau_{i}f(\ldots,z_{i},z_{i+1},\ldots)=f(\ldots,z_{i+1},z_{i},\ldots)$. 

Consider the state which is not invariant under the action of $e_{i}$. The $\pi$-th element of (\ref{qKZ-t-e}) 
is $t_{i}\Psi_{\pi}({\bf{z}})=-\tau\Psi_{\pi}({\bf{z}})$, or equivalently, 
\begin{eqnarray}
(qz_{i}-q^{-1}z_{i+1})\tau_{i}\Psi_{\pi}({\bf z})=(qz_{i+1}-q^{-1}z_{i})\Psi_{\pi}({\bf z}).
\end{eqnarray}
Since $\Psi_{\pi}({\bf z})$ is supposed to be a polynomial, $\Psi_{\pi}({\bf z})$ has a 
factor $(qz_{i}-q^{-1}z_{i+1})$. 

Let $\tau_{i,i+l}:=\tau_{i}\ldots\tau_{i+l-2}\tau_{i+l-1}\tau_{i+l-2}\ldots\tau_{i}$ be an exchange operator 
such that \linebreak
$\tau_{i,i+l} f(\ldots,z_{i},\ldots,z_{i+l},\ldots) =f(\ldots,z_{i+l},$ $\ldots,z_{i},\ldots)$.
Then, we have 
\begin{eqnarray}\label{exchange-long}
\check{P}(z_{i},\ldots,z_{i+l})\Psi=\tau_{i,i+l}\Psi
\end{eqnarray}
where 
\begin{eqnarray}
\check{P}(z_{i},\ldots,z_{i+l})&=&\check{R}_{i}(z_{i+l},z_{i+1})\ldots\check{R}_{i+l-2}(z_{i+l},z_{i+l-1})
\nonumber  \\
&&\cdot\check{R}_{i+l-1}(z_{i+l},z_{i})\check{R}_{i+l+2}(z_{i+l-2},z_{i})\ldots\check{R}_{i}(z_{i+1},z_{i}).
\end{eqnarray}
Consider a state $\pi$ which is not invariant under the action of $e_{i+j}$ for $0\le j\le l-1$. 
Taking into account that $\check{R}_{i+l-1}(z_{i+l},z_{i})\propto e_{i+l-1}$ if we set $z_{i+l}=q^{-2}z_{i}$,  
it is shown that
\begin{eqnarray}
\tau_{i,i+l}\Psi_{\pi}({\bf z})|_{z_{i+l}=q^{-2}z_{i}}=0.
\end{eqnarray}
This means $\Psi_{\pi}({\bf z})$ has a factor $(qz_{i}-q^{-1}z_{i+l})$. Therefore, in total  
$\Psi_{\pi}({\bf z})$ has factors $\prod_{i\le m<n\le i+l}(qz_{m}-q^{-1}z_{n})$. 

\paragraph{Highest weight state}
The highest state $\pi^{0}$ of this model is given by the following path:
\begin{eqnarray}
\scalebox{0.5}{\includegraphics{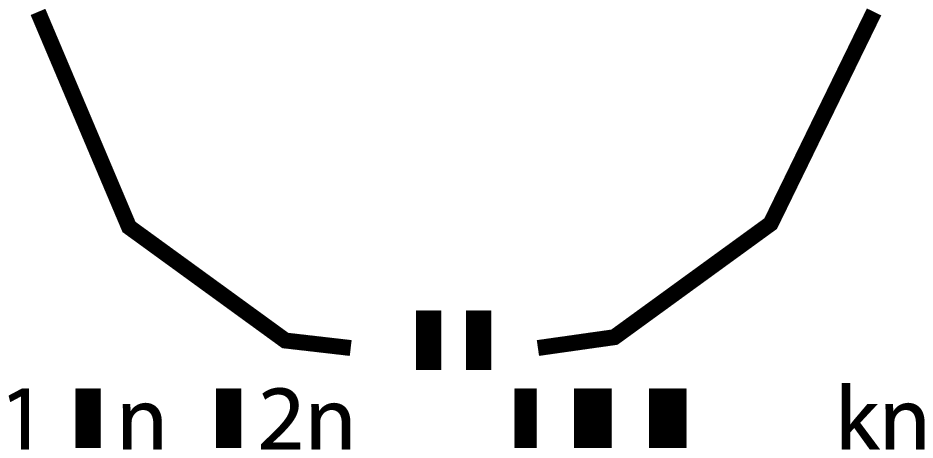}}
\end{eqnarray}
$\pi^{0}=\{\pi_{i}|\pi_{pn+q}=k+1-p, 0\le p\le k-1, 1\le q\le n \}$. This highest weight is characterized by 
\begin{eqnarray}
e_{kn}\pi^{0}&=&\tau \pi^{0},  \\
e_{i}\pi^{0}&\neq& \pi^{0},\ \ \ \mathrm{for}\ 1\le i\le kn-1.
\end{eqnarray}
The number $\sharp\{i |e_{i}\pi^{0}=\tau\pi^{0}\}$ is minimal. The highest state is invariant (up to a constant) 
only under the action of $e_{kn}$. The entry $\Psi_{\pi^{0}}$ is written as 
\begin{eqnarray}\label{Psi-highest}
\Psi_{\pi^{0}}=\prod_{1\le i<j\le nk}(qz_{i}-q^{-1}z_{j}), 
\end{eqnarray}
under the assumption of the minimal degree. 
The total degree of $\Psi_{\pi^{0}}$ is $N(N-1)/2$ and the partial degree is $N-1$ for each $z_{i}$. 

\paragraph{Recursive relation}
Let us fix an integer $m$ and take a special parameterization of the form 
\begin{eqnarray}\label{specialization-z}
z_{m+j}=q^{-2j}z, \qquad 0\le j\le k-1. 
\end{eqnarray}
This kind of specializations is called the wheel condition in the theory of symmetric polynomials~\cite{FJMM03}.
The entry $\Psi_{\pi}({\bf z})$ is non-vanishing only when $\pi$ has the convex 
sequence $\pi_{m+j}=j+1$ for $0\le j\le k-1$. 

For an unrestricted path $\pi$ of length $(n-1)k$, let $\varphi_{m,m+k-1}(\pi)$ be an embedded path of 
length $nk$ where the convex sequence of length $k$ is inserted between $\pi_{m}$ and $\pi_{m+1}$.
Then, we have the following recursion relation
\begin{eqnarray}\label{rr-psi}
\Psi_{\varphi_{m,m+k-1}(\pi)}({\bf z})|_{(\ref{specialization-z})}&=&C z^{\frac{1}{2}k(k-1)}
\left(\prod_{1\le j\le N \atop j\neq m,\ldots,m+k-1} (qz_{i}-q^{-1}z)^{k} \right)
\Psi_{\pi}({\bf z'})
\end{eqnarray}
where ${\bf z'}={\bf{z}}\backslash \{z_{m},\ldots,z_{m+k-1}\}$ and $C$ is some constant depending only 
$q$ and $N$. To see this relation, suppose that $\Psi({\bf z})$ is the minimal degree solution for $N$
variables. The r.h.s of (\ref{rr-psi}) satisfies the $q$-KZ equation for $N-k$ variables. This assures 
that $\Psi({\bf z'})$ is also the solution of the $q$-KZ equation. 
Note that the total degree and partial degrees with respect to all $z_{i}$ are consistent.

\paragraph{Razumov-Stroganov point and the sum rule}
We define the simultaneous eigen covector $v$ satisfying 
\begin{eqnarray}
ve_{i}&=&\tau v, \\
v\sigma&=&v,  
\end{eqnarray}
or we may write as $v\check{R}_{ii+1}=v$ for all $i$. The existence of $v$ requires that $q$ be a root of unity. 
Together with the vanishing condition of the $q$-symmetrizer (\ref{vc-Hecke}), we should have 
$U_{k}(\tau)=0$, \textit{i.e.}, we should take the Razumov-Stroganov (RS) point 
$q=-\exp\left(\frac{i\pi}{k+1}\right)$. In the below, $q$ is taken as this RS point. 

The sum rule is the formula for the weighted sum, 
$W(\mathbf{z})=v\cdot \Psi=\sum_{\pi}v_{\pi}\Psi_{\pi}(\mathbf{z})$. 
One can show that $W(\mathbf{z})$ is a homogeneous and symmetric polynomial with respect to all the variables $z_{i}$, 
which is led from the fact that the polynomial $\Psi_{\pi^{0}}({\bf z})$ is homogeneous and the actions of $t_{i}$ 
preserve this property. Since $\tau_{i}W(\mathbf{z})=v\cdot \tau_{i}\Psi=v\cdot\check{R}_{i}\Psi=W(\mathbf{z})$, 
$W(\mathbf{z})$ is a symmetric polynomial. From the recursive relation (\ref{rr-psi}), we have the recursive relation
\begin{eqnarray}
W(\mathbf{z})|_{(\ref{specialization-z})}=C z^{\frac{1}{2}k(k-1)}
\left(\prod_{1\le j\le N \atop j\neq m,\ldots,m+k-1} (qz_{i}-q^{-1}z)^{k} \right)
W({\bf z'}).
\end{eqnarray}
The total degree and partial degree of $W({\bf z})$ are $N\choose 2$ and $N-1$ respectively. 
From the above observation and Proposition~\ref{prop-rrschur}, we can show that the sum $W(\mathbf{z})$ is written in 
terms of Schur functions $s_{\lambda}({\bf z})$ as
\begin{eqnarray}
W(\mathbf{z})=(\mathrm{const.})\prod_{l=0}^{k-1}s_{Y_{k,l}^{n}}(z_{1},\ldots,z_{nk})
\end{eqnarray}
with an appropriate overall normalization. Here, the Young diagrams are $Y_{k,l}^{n}:=\delta(n^{l},n-1^{k-l})$ with  
\begin{eqnarray}
\delta(n^{l},n-1^{k-l})
=(\underbrace{n,\cdots,n}_{l},\underbrace{n-1,\cdots,n-1}_{k},\underbrace{n-2,\cdots,n-2}_{k}
,\cdots,\underbrace{1,\cdots,1}_{k}).
\end{eqnarray}

\noindent
{\bf Remark:}
When $k=2$, we have 
\begin{eqnarray}
W(z_{1},\cdots,z_{2n})=(\mathrm{const.})s_{Y_{2,0}^{n}}(z_{1},\cdots,z_{2n})s_{Y_{2,1}^{n}}(z_{1},\cdots,z_{2n}).
\end{eqnarray}
We reproduce the sum for the $O(1)$ loop model on a cylinder~\cite{DiFZJZu06,KasPas06}.  

\subsection{Relation to special solutions of the $q$-KZ equation}\label{subsec-specialqKZ}
In this subsection, we show that the eigenvector of the transfer matrix of the $A_{k}$ generalized model at 
the Razumov-Stroganov point is viewed as the special solution of the $q$-KZ equation 
at $(q,s)=(-\exp(i\pi/(k+1)),1)$ in~\cite{KasTak06a}. 

In Section 4 of~\cite{KasTak06a}, special solutions of the $q$-KZ equation were constructed from 
the non-symmetric Macdonald polynomial~\cite{Mac03} through the action of the affine Hecke algebra. 
Consider the $q$-KZ equation on the spin representation instead of on the space of paths. 
Let $l$ and $r$ be positive integers such that $1\le l\le \min\{N-1,k\}$, $r\ge 2$ and $l+1$ and $r-1$ are coprime. 
We take the specialization 
\begin{eqnarray}\label{specialization-qs}
q^{2(l+1)}s^{-(r-1)}=1.
\end{eqnarray}
An element $\lambda=(\lambda_{1},\ldots,\lambda_{N})\in\mathbb{Z}^{N}$ is said to be admissible if $\lambda$ satisfies 
\begin{eqnarray}
&&\lambda_{i}^{+}-\lambda_{i+l}^{+}\le r-1\qquad \mbox{for any $1\le i\le n-l$},\\
&&\lambda_{i}^{+}-\lambda_{i+l}^{+}=r-1\qquad \mbox{only if $w_{\lambda}^{+}(i)<w_{\lambda}^{+}(i+k)$}.
\end{eqnarray}
Here, $\lambda^{+}$ is the unique dominant in $\mathfrak{S}_{N}\lambda$, 
{\it i.e.} $\lambda^{+}_{1}\ge\lambda^{+}_{2}\ge\ldots\ge\lambda^{+}_{N}$, 
and $w_{\lambda}^{+}$ is the shortest element in $\mathfrak{S}_{N}$ 
such that $w_{\lambda}^{+}\lambda^{+}=\lambda$. 
Let $\mu\in\mathbb{Z}^{N}$ be an element constructed from an dominant element $a\in\mathbb{Z}^{l}$ (see Section 4.3 
in~\cite{KasTak06a}).  
Then, the solution of the $q$-KZ equation of level $\frac{l+1}{r-1}-N$ is created from the non-symmetric 
Macdonald polynomial $E_{\mu}$ with the specialization (\ref{specialization-qs}) and an 
admissible $\mu$ (lemma 4.5 and Theorem 4.6 in~\cite{KasTak06a}).

We want to find the solution of the $q$-KZ equation on the spin basis, which is dual to the eigenvector of 
the $A_{k}$ generalized model. The eigenvector of the transfer matrix of the $A_{k}$ generalized model 
at the Razumov-Stroganov point is characterized by the highest weight state (\ref{Psi-highest}). A 
monomial $\prod_{j=1}^{N}z_{j}^{N-j}$ is the dominant one in the expansion of the r.h.s. of 
Eqn.(\ref{Psi-highest}). Once the higest weight state is fixed, the eigenvector of the $A_{k}$ 
generalized model is uniquely determined.  

We restrict ourselves to the special solution of the $q$-KZ equation with the level $1-k+\frac{1}{k}$ 
on the spin basis at the specializations
\begin{eqnarray}
&&(l,r)=(k,k+1),\label{special-lr}\\
&&a=(nk-1,\ldots,k(n-1))\in\mathbb{Z}^{k}, \label{special-a} \\
&&(q,s)=(-\exp(i\pi/(k+1)),1). \label{special-qsRS}
\end{eqnarray}
Note that the dominant monomial of the solution is $\prod_{j=1}^{N}z_{j}^{N-j}$. 

From the construction, a state of the $A_{k}$ generalized model is written as a linear combination of the 
spin basis (see Section~\ref{states-Ak} and~\ref{subsec-Ak-spin}). Therefore, a suitable linear comibination of the 
above solution of the $q$-KZ equation on the spin basis gives the eigenvector of the transfer matrix of the 
$A_{k}$ generalized model at the RS point. The eigenvector of the $A_{k}$ generalized model and the special 
solution of the $q$-KZ equation share the same dominant monomial. Together with the uniqueness of 
the eigenvector of the $A_{k}$ generalized model, the eigenvector of the $A_{k}$ generalized model at the 
Razumov-Stroganov point coincides with the special solution of the $q$-KZ equation characterized by 
Eqns.(\ref{special-lr})-(\ref{special-qsRS}). When $k=2$, we can see the solution with $(l,r)=(2,3)$ as in~\cite{KasPas06}.

\section{Recursive Relation for Schur Functions}\label{sec-schur}
Let us denote Young diagrams by $Y_{k,l}^{n}=\delta(n^{l},n-1^{k-l})$ and $Y_{k,l}^{1}=\delta(1^{l},0^{k-l})$
where 
\begin{eqnarray}
\delta(n^{l},n-1^{k-l})
=(\underbrace{n,\cdots,n}_{l},\underbrace{n-1,\cdots,n-1}_{k},\underbrace{n-2,\cdots,n-2}_{k}
,\cdots,\underbrace{1,\cdots,1}_{k}).
\end{eqnarray}

\begin{proposition}\label{pr-spe-Y1}
When $q=-\exp(\frac{i\pi}{k+1})$, the principal specialization of $s_{Y_{k,l}^{1}}$ is given by 
\begin{eqnarray}
s_{Y^{1}_{k,l}}(1,q^{2},\cdots,q^{2(k-1)})=(-1)^{l}q^{-2l}.
\end{eqnarray}
\end{proposition}

\begin{proposition}\label{prop-rrschur}
Set $q=-\exp(\frac{i\pi}{k+1})$. The following recursion relation for the Schur 
function $s_{Y_{k,l}^{n}}$ holds:
\begin{eqnarray}
s_{Y_{k,l}^{n}}(z_{1},\cdots,z_{nk})\big|_{wheel}
&=&(-)^{l}q^{-2l} z^{l}\prod_{i=1}^{(n-1)k}(z_{i}-q^{2k}z) s_{Y_{k,l}^{n-1}}(z_{1},\cdots,z_{k(n-1)})
\label{rrschur}
\end{eqnarray}
where the wheel condition (the specialization of variables in Eqn.(\ref{rrschur}), see also~\cite{FJMM03}) 
is such that $z_{j_{m}}=q^{2(j_{m}-1)}z$ for $1\le m\le k$ and $j_{m}<j_{m+1}$. 

Further, if we write as $S^{n}({\mathbf{z}}):=\prod_{l=0}^{k-1}s_{Y_{k,l}^{n}}$, we have the following 
recursive relation from the proposition~\ref{prop-rrschur},
\begin{eqnarray}
S^{n}({\mathbf z})|_{z_{k(n-1)+j}=q^{2(j-1)}z}=(-)^{k(k+1)/2}q^{2}
z^{k(k-1)/2}\prod_{i=1}^{(n-1)k}(z_{i}-q^{2k}z)^{k}S^{n-1}({\mathbf z'})
\end{eqnarray}
where ${\mathbf z'}={\mathbf z}\backslash\{z_{k(n-1)+1},\cdots,z_{nk}\}$. 
\end{proposition}
\begin{proof}
Since a Schur function $s_{\lambda}(x)$ is a symmetric function with respect to all the 
variables $x_{i}$, we consider only the following specialization of variables without loss of 
generality:
\begin{eqnarray}\label{special-z}
z_{k(n-1)+j}=q^{2(j-1)}z, \qquad 1\le j\le k.
\end{eqnarray}

Since the number of boxes in the first column of the Young diagram $Y_{k,l}^{n}$ is 
$k(n-1)+l$, we have at least a factor $z^{l}$ when we take the specialization (\ref{special-z}). 
Around $z=0$, the l.h.s of  (\ref{rrschur}) is approximated as 
\begin{eqnarray}
\begin{aligned}
s_{Y_{k,l}^{n}}(z_{1},\cdots,z_{nk})\big|_{z_{k(n-1)+j}=q^{2(j-1)}z}
&\propto z^{l}\prod_{i=1}^{(n-1)k}z_{i}\cdot s_{Y_{k,l}^{n-1}}(z_{1},\cdots,z_{k(n-1)})+O(z^{l+1}).
\end{aligned}
\label{schurz}
\end{eqnarray}
The maximal and minimal degrees with respect to $z$ of the first term of the r.h.s. of Eqn.(\ref{schurz})  
is $k(n-1)+l$ and $l$ respectively.

A Schur function of $m$-variables has an expression in terms of determinants as
\begin{eqnarray}
s_{\lambda}(z_{1},\cdots,z_{m})=\frac{\det z_{i}^{\lambda_{j}+m-j}}{\det z_{i}^{m-j}}
=\frac{\det z_{i}^{\lambda_{j}+m-j+l+1}}{\prod z_{i}^{l+1}\det z_{i}^{m-j}}.
\label{schurdef}
\end{eqnarray}
For $\lambda=Y_{k,l}^{n}$ and $m=nk$, the sequence $\lambda^{'}_{j}=\lambda_{j}+m-j+l+1$ 
has the form 
\begin{eqnarray}
\lambda'=(l,\cdots,1,\underbrace{k,k-1,\cdots,1}\cdots,\underbrace{k,\cdots,1})\ \ \mathrm{mod}\ k+1. 
\end{eqnarray}
Note that there is all $\lambda'_{j}$ is not a multiple of $(k+1)$.   

Specialize $k$ variables as $z_{k(n-1)+j}=q^{2(j-1)}z\ (1\le j\le k)$ and set $z_{i}=q^{2k}z$ for 
$i\neq k(n-1)+j, 1\le\forall j\le k$ in the determinant expression (\ref{schurdef}). 
We find that $k+1$ row-vectors in the $\det$ of the numerator of (\ref{schurdef}), say $v_{r}$, $0\le r\le k$,  
are such that $(v_{r})_{i}=q^{2r\lambda_{i}^{'}}z^{\lambda_{i}^{'}}$. These $k+1$ row-vectors are not 
linearly independent, since we have
\begin{eqnarray}
\sum_{r=0}^{k}(v_{r})_{i}=\left(\sum_{r=0}^{k}q^{2r}\right)q^{\lambda_{i}^{'}}z^{\lambda_{i}^{'}}=0
\end{eqnarray}
where we have used the relation $\sum_{r=0}^{k}q^{2r}=0$ for $q=-\exp(\pi i/(k+1))$. 
The determinant turns to be zero under the above specialization of $k+1$ variables. 
By the symmetry of the variables in the Schur function, we find that the l.h.s of (\ref{rrschur}) has 
a factor $\prod_{1\le i\le n(k-1)}(z_{i}-q^{2k}z)$. Together with (\ref{schurz}), we may write as
\begin{eqnarray}
s_{Y_{k,l}^{n}}(z_{1},\cdots,z_{nk})\big|_{z_{k(n-1)+j}=q^{2(j-1)}z}
\propto z^{l}\prod_{i=1}^{(n-1)k}(z_{i}-q^{2k}z)\cdot s_{Y_{k,l}^{n-1}}(z_{1},\cdots,z_{k(n-1)}).
\label{schurzz}
\end{eqnarray}
Actually, the total degree and partial degree of $z_{i}$ in the both sides of (\ref{schurzz}) coincide. 
The prefactor $(-)^{l}q^{-2l}=s_{Y^{1}_{k,l}}(1,q^{2},\cdots,q^{2(k-1)})$ is checked by collecting the 
terms with the lowest degree in $z$ (see Proposition~\ref{pr-spe-Y1}).
\end{proof}

\section{Conclusion}\label{sec-conclusion}
In this paper we have defined and studied the $A_{k}$ generalized model of the $O(1)$ loop model on a cylinder by using the 
representation of the affine Hecke algebra. The affine Hecke algebra is characterized by extra novel
vanishing conditions, the cylindric relations. Two representations of the algebra have been given; the
first one is based on the spin representation, and the other is based on states of the $A_{k}$ generalized model. 
These two representation are connected by the word representation of states of the $A_{k}$ generalized model. 
We have established an explicit way of constructing states of the $A_{k}$ generalized model by the use 
of the rhombus tiling. We have shown that the Yang-Baxter equation and $q$-symmetrizers are depicted 
as hexagons and polygons, respectively.
The meaning of the cylindric relations is clearly seen in 
the graphical depiction of the $A_{k}$ generalized model. The cylindric relations for the affine 
Temperley-Lieb algebra implies that a loop 
surrounding the cylinder returns a weight $\tau$. For the $A_{k}$ generalized model, a ``band" consisting of rhombi surrounding 
the cylinder gives a certain weight in terms of the second kind of Chebyshev polynomials. 

We have considered the eingenvector of the transfer matrix of the $A_{k}$ generalized model at the Razumov-Stroganov 
point, $q=-\exp(\pi i/(k+1))$. 
It has been found that this eigenvector coincides with the special solution of the $q$-KZ equation of 
level $1+\frac{1}{k}-k$ at $q=-\exp(\pi i/(k+1))$ and $s=1$.
We have examined the sum rule for the $A_{k}$ generalized model on a cylinder and shown the formula 
is written in terms of the product of $k$ Schur functions. The obtained sum rule 
includes the sum rule for the $O(1)$ loop model on a cylinder when $k=2$.

There are still some open problems. It was shown that the sum rules for the $O(1)$ loop models with various 
boundary conditions are related to exactly solvable models with symmetries, alternating sign matrices 
with certain symmetries, or total numbers 
of the plane partitions with symmetries. We expect that the sum rule for the $A_{k}$ generalized model 
on a cylinder may also relate to those objects. 
In the case of $k=2$, the total number of half-turn symmetric alternating sign matrices appeared 
in this context~\cite{DiFZJZu06}. 
The method used in this paper is applicable to the Hecke algebras of other types. 
We hope to come back to these issues in the future.  

\paragraph*{Acknowledgement}
The authors express thanks to Professor Miki Wadati for critical reading of the manuscript and continuous 
encouragements.

\bigskip
\appendix
\renewcommand{\thesection}{Appendix~\Alph{section}}
\section{}
\renewcommand{\thesection}{\Alph{section}}
\subsection{}\label{Cipipi'}
In this appendix, we will show that a class of $C_{i,\pi,\pi^{'}}$ (see (P\ref{P1}-\ref{P4}) in Section~\ref{sec-Akmodel}) is 
equal to $1$. Let us recall that the coefficient $C_{i,\pi,\pi^{'}}$ may be non-zero when $\pi_{i}\le \pi_{i+1}$. 

We introduce a sequence of $\check{L}$-matrices as 
\begin{eqnarray}
\mathcal{L}_{i+1,i+l}(m)=\check{L}_{i+1}(m)\check{L}_{i+2}(m+1)\cdots\check{L}_{i+l}(m+l-1),
\end{eqnarray}
and $\mathcal{L}_{i,j}=1$ if $i>j$. Let $\mathcal{B}$ be a word representation corresponding a 
state $|\pi\rangle$ with $\pi_{i+l+1}<\pi_{i}<\pi_{i+1}<\ldots<\pi_{i+l}$ for $l\ge1$. We consider a word of 
the form, $\mathcal{B}^{'}:=\mathcal{L}_{i+1,i+l}(m^{'})\mathcal{B}$. 
\begin{proposition}
The action of $\check{L}_{i}(m)$ on $\mathcal{B}^{'}$ is given by
\begin{eqnarray}\label{R-B-recursion}
\check{L}_{i}(m)\mathcal{B}^{'}=\mathcal{L}_{i,i+l}(m)\mathcal{B}
+\mathcal{L}_{i+2,i+l}(m)\mathcal{B}.
\end{eqnarray}
\end{proposition}
\begin{proof}
We use the method of induction. we assume
\begin{eqnarray}
\check{L}_{i+l'-1}(m)\mathcal{L}_{i+l',i+l}(m)\mathcal{B}=\mathcal{L}_{i+l'-1,i+l}(m)\mathcal{B}
+\mathcal{L}_{i+l'+1,i+l}(m)\mathcal{B}
\end{eqnarray}
for $1\le l^{'}\le l-1$. From the above assumption, we have 
\begin{eqnarray}
\check{L}_{i}(m)\mathcal{B}^{'}&=&\check{L}_{i}(m)\mathcal{L}_{i+1,i+l-1}(m)
\check{L}_{i+l}(m+l-1)\mathcal{B} \nonumber \\
&=&(\mathcal{L}_{i,i+l-1}(m)+ \mathcal{L}_{i+2,i+l-1}(m)\check{L}_{i+l}(m+l-1)\mathcal{B} \nonumber \\
&=&\mathcal{L}_{i,i+l}(m)\mathcal{B}+\mathcal{L}_{i+2,i+l}(m)\mathcal{B} \nonumber  \\
&&+(\Delta_{m+l-1}\mathcal{L}_{i,i+l-1}(m)-\Delta_{m+l-2}\mathcal{L}_{i+2,i+l-1}(m))\mathcal{B} 
\label{R-B-recursion2}
\end{eqnarray}
where we used the word $\check{L}_{i+l}(m+l-1)\mathcal{B}$ satisfies 
$\pi_{i+l}<\pi_{1}<\ldots<\pi_{i+l-1}$ and $\Delta_{k}=\mu_{k}-\mu_{k-1}$. By using $e_{j}\mathcal{B}=\tau\mathcal{B}$ 
for $i\le j\le i+l-1$, the third term in (\ref{R-B-recursion2}) is 
\begin{eqnarray}
\text{the 3-rd term}&=&
\prod_{r=1}^{l-2}\frac{1}{\mu_{m+r-1}}\left(\frac{\Delta_{m+l-1}}{\mu_{m+l-2}\mu_{m+l-1}}-\Delta_{m+l-2}\right)
\mathcal{B}  \nonumber \\
&=&0
\end{eqnarray}
where we used the relation $\frac{1}{\mu_{k}\mu_{k-1}}\Delta_{k}=\Delta_{k-1}$.
We finally obtain Eqn. (\ref{R-B-recursion}) holds true by induction.
\end{proof}

Set $m=1$ in Eqn. (\ref{R-B-recursion}). Together with the construction of states considered in 
Section~\ref{subsec-Akmodel}, we have the following corollary:
\begin{corollary}
Suppose that a state $|\pi^{0}\rangle$ is equivalent to $\mathcal{B}'$ in the word representation and 
$\pi^{0}_{i+1}\le\pi_{i-1}^{0}$. We have 
\begin{eqnarray}
e_{i}|\pi^{0}\rangle=|\pi^{1}\rangle+|\pi^{2}\rangle
\end{eqnarray}
where
\begin{eqnarray}
\pi^{1}=\left\{
\begin{split}
&\pi^{1}_{i}=\pi_{i+1}^0, \pi^1_{i+1}=\pi_{i}^{0}, \\ 
&\pi^{1}_{j}=\pi^{0}_{j}\quad \mbox{for}\quad j\neq i,i+1,
\end{split}\right. \qquad
\pi^{2}=\left\{
\begin{split}
&\pi^{2}_{i+1}=\pi^{0}_{i+2}, \pi^{2}_{i+2}=\pi^{0}_{i+1},\\
&\pi^{2}_{j}=\pi^{0}_{j}\quad \mbox{for}\quad j\neq i+1,i+2.
\end{split}\right.
\end{eqnarray}
\end{corollary}

\subsection{}\label{examples-rep}
\subsubsection{}
We have six states in the case of $k=N=3$. The word representation of states is listed as 
 \begin{center}
  \begin{tabular}{c|cccccc}
word &  $e_{3}Y_{2}$     &  \ \  $Z_{2,3}Y_{2}$ & \ \ $Z_{13}Y_{2}$   & \ \ $e_{1}Z_{2,3}Y_{2}$\ \ 
   &  $e_{2}Z_{1,3}Y_{2}$ \ \    &   $Y_{2}$      \\ \hline
  path  & $321$   &  $312$  &  $231$  &  $132$  & $213$   & $123$   \\
  \end{tabular}
 \end{center}
where $Z_{i,j}=e_{i}e_{j}-1$ and $Y_{2}=e_{1}e_{2}e_{1}-e_{1}$. We obtain the representation 
of the generators: 
\begin{center}
$e_{1}=$\scalebox{0.8}{$\left(
  \begin{array}{cccccc}
     0  &  0  &  0  &  0  &  0  &  0  \\
     0  &  0  &  0  &  0  &  0  &  0  \\
   \tau^{2}-1 & 0  & \tau   & 0  & 1   &  0  \\
     0  &   1  & 0   & \tau   &  0  & 0  \\
     0  &  0  &  0  & 0   &  0  &  0  \\
     1  &  0  &  0  & 0   & \tau^{2}-1   & \tau   \\
  \end{array}
\right)$},\ \ 
$e_{2}=$\scalebox{0.8}{$\left(
  \begin{array}{cccccc}
     0  &  0  &  0  &  0  &  0  &  0  \\
   \tau^{2}-1 & \tau & 0   & 1  & 0   &  0  \\
     0  &  0  &  0  &  0  &  0  &  0  \\
     0  &  0  &  0  & 0   &  0  &  0  \\
     0  &  0  &  1  & 0   & \tau   &  0  \\
     1  &  0  &  0  & \tau^{2}-1 & 0  & \tau  \\
  \end{array}
\right)$},  \\
$e_{3}=$\scalebox{0.8}{$\left(
  \begin{array}{cccccc}
   \tau & 0  &  0  & 0  & 0   &  1  \\
     0  &  \tau  &  0  & 1   & \tau^{2}-1   &  0  \\
     0  &  0  &  \tau  &  \tau^{2}-1 & 1   & 0  \\
     0  &  0  &  0  &  0  &  0  &  0  \\
     0  &  0  &  0  &  0  &  0  &  0  \\
     0  &  0  &  0  & 0   &  0  &  0  \\
  \end{array}
\right)$}, \ \ 
$\sigma=$\scalebox{0.8}{$\left(
  \begin{array}{cccccc}
     0 & 0  &  0  & 1  & 0   &  0  \\
     0  &  0  &  1  & 0  & 0  &  0  \\
     0  &  0  &  0  &  0 & 0   & 1 \\
     0  &  0  &  0  &  0  &  1  &  0  \\
     1  &  0  &  0  &  0  &  0  &  0  \\
     0  &  1  &  0  & 0   &  0  &  0  \\
  \end{array}
\right)$}.
\end{center}
\subsubsection{}
Two examples how the generator of the affine Hecke algebra acts on a state. We consider 
the case where $k=4$ and $n=2$. 
\begin{eqnarray}
\scalebox{2}{$e_{4}$} \raisebox{-0.4\height}{\scalebox{1.0}{\includegraphics{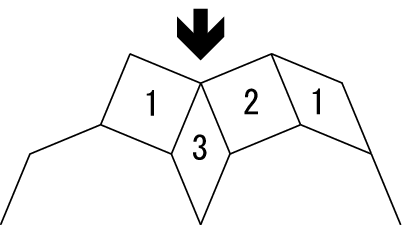}}}&=&
\raisebox{-0.4\height}{\scalebox{1.0}{\includegraphics{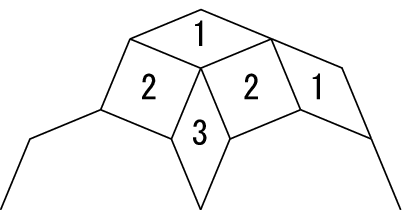}}}+
\raisebox{-0.4\height}{\scalebox{0.8}{\includegraphics{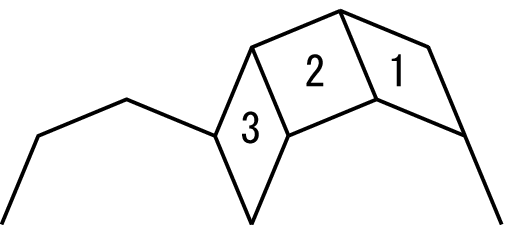}}} \label{app-example}\\
\scalebox{2}{$e_{5}$} \raisebox{-0.4\height}{\scalebox{1.0}{\includegraphics{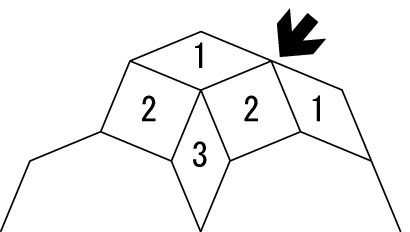}}}&=&
\raisebox{-0.4\height}{\scalebox{1.0}{\includegraphics{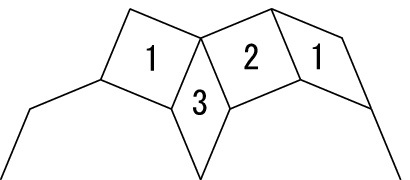}}}+
\raisebox{-0.4\height}{\scalebox{0.4}{\includegraphics{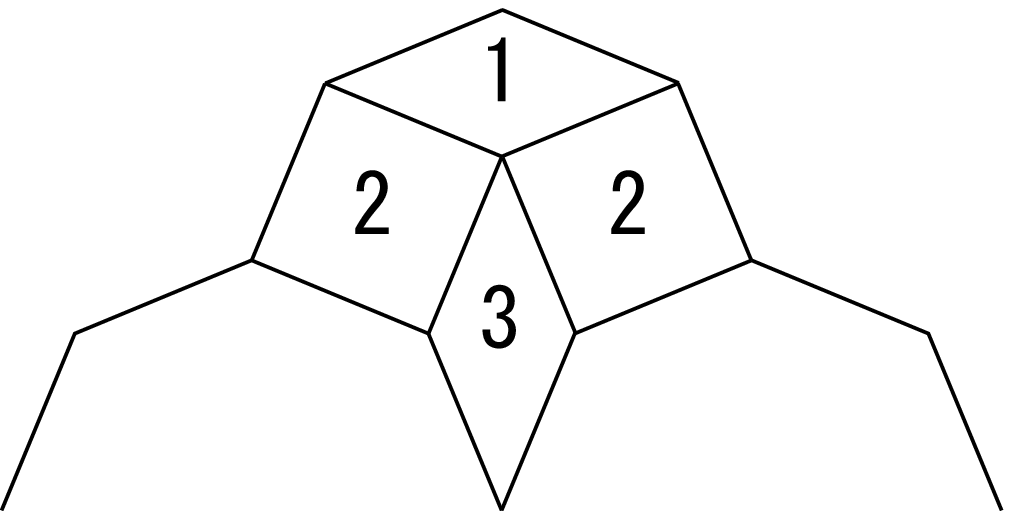}}}
\end{eqnarray}
The bold arrows indicate where the generators act.  
The state of the l.h.s. of Eqn. (\ref{app-example}) is an example of states which do not satisfy the 
zero-sum rule on the top vertices of $Y_{q\mbox{-sym}}$. We see that the properties (P\ref{P2}) and 
(P\ref{P4}) are satisfied.

\end{document}